\documentclass[preprint,12pt]{elsarticle}

\usepackage{graphicx}
\usepackage{dcolumn}
\usepackage{bm}
\usepackage{amssymb}
\usepackage{natbib}
\usepackage{color}
\biboptions{sort&compress}

\journal{Mat. Met. in Medical and Biological Sciences}

\begin{document}

\begin{frontmatter}

\title{Fractional and fractal extensions of epidemiological models}

\author{Enrique C. Gabrick$^{1}$, Ervin K. Lenzi$^{1,2}$, Antonio M.
Batista$^{1,3}$}
\address{$^1$Graduate Program in Science, State University of Ponta Grossa, 
84030-900, Ponta Grossa, PR, Brazil.}
\address{$^2$Department of Physics, State University of Ponta Grossa, 
84030-900, Ponta Grossa, PR, Brazil.}
\address{$^3$Department of Mathematics and Statistics, State University of
Ponta Grossa, 84030-900, Ponta Grossa, PR, Brazil}

\cortext[cor]{ecgabrick@gmail.com (ECG); eklenzi@uepg.br (EKL); 
antoniomarcosbatista@gmail.com (AMB).}

\begin{abstract}
The spread of disease is a challenging problem for public health. One way to study the spread of disease is through mathematical models. The most successful models compartmentalize the host population according to their infectious stage, e.g., susceptible (S), infected (I), exposed (E), and recovered (R). The composition of these compartments leads to the SI, SIS, SIR, and SEIR models, which employ standard ordinary differential equations. In this Chapter, we present and compare three formulations of SI, SIS, SIR, and SEIR models in the framework of standard (integer operators), fractional (Caputo sense), and fractal derivatives (Hausdorff sense). Firstly, we introduce basic concepts of fractional operators, such as the numerical integration via the Euler method and their convergence. Following, we present the fractal derivative and its connection with integer operators. After, we introduce the models and applications in real data and analyze the differences between each formulation. As an application of the SI model, we study the evolution of AIDS cases in Bangladesh from 2001 to 2021. For this case, our simulations suggest that fractal formulation describes the data well. For the SIS model, we consider syphilis data from Brazil from 2006 to 2017. In this case, the three frameworks describe the data with good accuracy. We used data from Influenza A to adjust the SIR model in previous approaches and observed that the fractional formulation was better. The last application considers the COVID-19  data from India in the range 2020-04-10 to 2020-12-31 to adjust the parameters of the SEIR model. The standard formulation fits the data better than the other approaches. As a common result, all models exhibit steady solutions in the different formulations. The time to reach a steady solution is correlated to the considered approach. The standard and fractal formulations reach the steady state earlier when compared with the fractional formulation.  
\end{abstract}

\begin{keyword}
Epidemiology \sep Infectious diseases \sep Fractional calculus \sep Fractal calculus \sep SI model \sep SIS model \sep SIR model \sep SEIR model \sep AIDS \sep Syphilis \sep Influenza \sep COVID-19
\end{keyword}

\end{frontmatter}

%%%%%%%%%%%%%%%%%%%%%%%%%%%%%%%%%%%%%%%%%%%%%%%%%%%
%%%%%%%%%%%%%%%%%%%%%%%%%%%%%%%%%%%%%%%%%%%%%%%%%%%
\section{Introduction} \label{Introduction}

Infectious diseases have affected populations throughout history and, in general, caused many deaths before disappearing. For example, the  Black Death killed more than 25 million people in Europe~\cite{Glatter2021}, the Spanish flu (1918-1919) caused around 20 to 50 million deaths people worldwide~\cite{Tumpey2005}, and, more recently, COVID-19 generated about 7 million of death~\cite{WHO-covid19}. Moreover, diseases affect not only people but also plants~\cite{Sankaran2010} and animals~\cite{Galvis2022, Machado2022, Favero2017}. Therefore, disease spread is a crucial problem for socioeconomic development \cite{SinghBookb, SinghBookc}.

In the sense of understanding, forecasting, and studying control measures for disease spread, mathematical models are a powerful tool~\cite{Huppert2013, Wearing2005, Viboud2006, Grenfell2001, Keeling2005, Keeling2001, Hethcote1989, Hethcote2000}. The first model was proposed in 1662 by John Graunt (1620-1674) in his book: ``Natural and Political Observations Made upon the Bills of Mortality". Graunt analyzed Bills of Mortality records and gave a method to estimate the comparatives of being killed by different diseases. Daniel Bernoulli (1700–1782) proposed the next famous model in his works~\cite{Bernoulli1760}. Bernoulli's model was proposed to study smallpox dissemination. A good review of this model is found in Ref.~\cite{Dietz2002}. In 1855, another notable model was studied by John Snow (1813-1858) about the cholera cases in London~\cite{Snow1855}. A few years later, in 1873, William Budd (1811-1880) analyzed the typhoid spread~\cite{Budd1918}. The modern models are mainly based on the works developed during the years 1900 to 1935, especially by Sir R.A. Ross (1857-1932), W.H. Hamer, A.G. McKendrick (1876-1943), and W.O. Kermack (1898-1970)~\cite{Castillo-Chavez}. In particular, the recent epidemiological models are based on the works of McKendrick and  Kermack that were published in 1927~\cite{McKendrick1927}, 1932~\cite{McKendrick1932} and 1933~\cite{McKendrick1933}. These models compartmentalize the host population according to the status 
about the infectious stages. In this framework, it is possible to include new compartments to simulate: hospitalization~\cite{Dai2021}, asymptomatic/oligosymptomatic~\cite{Amaku2021}, quarantine~\cite{Safi2011}, two vaccination doses~\cite{Gabrick2022}, relaxation of control measures~\cite{Mugnaine2022}, easing restrictions~\cite{Silvio2021} and others~\cite{Steindorf2022, Wei2021, Marquioni2021}. Furthermore, epidemic models based on compartments have been showing excellent accuracy in reproducing real data~\cite{Manchein2020, Brugnago2020, Keeling2002, Olsen1990, Reyna-Lara2022, Soriano2020, Stollenwerk2015, Cooper2020}.

Despite the success of these models in reproducing real data based on integer differential equations, extensions for non-integer operators have been showing an improvement in some cases~\cite{Diethelm2013, Singh2020, Pinto2013, Chen2021, Area2015, Khan2021, Hassani2022, Khan2020, Pinto2022, Singh2021, Singh2021b}. These formulations are mainly based only on fractional operators~\cite{Evangelista2018, Evangelista2023, Singh2021c, Singh2022, Singh2020b}. Compositions of fractal and fractional operators also have been considered in extensions of epidemiological models~\cite{Sabbar2023, Cui2021, Addai2022, Liu2022, Qureshi2020, Ali2021, Abbasi2020, Mansour2021, Jian2022}. The literature evaluating only fractal operators in epidemiology is scarce \cite{Gabrick2023c}. 

This Chapter presents and discusses the models SI, SIS, SIR, and SEIR in their standard, fractional, and fractal formulation. By standard formulation, we refer to ordinary operators with integer order. We refer to Caputo sense and Hausdorff derivative by fractional and fractal formulation. We report essential dynamical aspects of the models and how they change in new formulations. Our simulations show that the main difference is in time to reach the steady solution, which takes more time in fractal one. In addition, the beginning and ending of the spread are described by different rates in different formulations. To exemplify the models, we show applications in real data. Considering the SI formulation, we adjust the AIDS data, which fractal operators describe well. For the SIS formulation, we consider syphilis data. In this case, the data is well-adjusted by the three formulations. We utilize the influenza data for the SIR model, which was better described by the fractional approach. Our last application is the description of COVID-19 data by the SEIR model. In this case, the three formulations describe the data with great accuracy. 

This Chapter is organized as follows: In Section~\ref{Non-integer}, we introduce the basic concepts about fractional operators (Subsection~\ref{Fractional_scheme}) and fractal operators (Subsection 2.2). The basic concepts of epidemiology are introduced in Section 3, followed by the presentation of SI (Subsection 3.1), SIS (Subsection 3.2), SIR (Subsection 3.3), and SEIR (Subsection 3.4) models. Finally, we draw our conclusions in Section 4.

%%%%%%%%%%%%%%%%%%%%%%%%%%%%%%%%%%%%%%%%%%%%%%%%%%%
\section{Non-integer calculus} \label{Non-integer}

The study of differential operators with non-integer order started with a letter from L'Hospital to Leibniz in 1695 asking him about the meaning of $d^\nu y /dx^\nu$ when $\nu = 1/2$~\cite{Evangelista2018}. After that, many researchers devoted themselves to understanding this problem, e.g., Euler, Lagrange, Laplace, Fourier, and others \cite{Guo2015}, giving rise to the fractional calculus \cite{Herrmann2014}. Although the start of this problem is not recent, many researchers have dedicated 
their work to fractional calculus development~\cite{Tenreiro2011, Caputo2008, Caputo2015, Ortigueira2015, Baleanu2018, Diethelm2002, Gorenflo1997}. In this context, many applications of fractional calculus have been made, such as electrical spectroscopy impedance~\cite{Rosseto2022,Scarfone2022,Lenzi2021}, wave propagation in porous media~\cite{cai2018survey}, quantum mechanics~\cite{Gabrick2023a, Gabrick2023b}, fluid mechanics~\cite{Kulish2002} and epidemiology~\cite{Chen2021}. Another possibility for non-integer calculus is fractal derivatives, as discussed in Refs.~\cite{Arif2021, Chen2010, Brouers2018, He2014, He2018} or combinations of fractal-fractional operators~\cite{Ali2021b, Qureshi2020, Atangana2017}. In this work, we aim to study the effects of non-integer order in epidemic models; before that, we introduce some basic concepts of fractional and fractal calculus. References more specialized in the subject are found in Refs.~\cite{Evangelista2018, Herrmann2014, Evangelista2023, SinghBook}.

%%%%%%%%%%%%%%%%%%%%%%%%%%%%%%%%%%%%%%%%%%%%%%%%%%%
\subsection{Fractional operators}\label{Fractional_scheme}

In this subsection, we introduce some basic concepts of fractional calculus \cite{Evangelista2018, Guo2015, Herrmann2014, Evangelista2023}. As the epidemiological equations are non-linear, analytical solutions are very hard or impossible to obtain in most part of cases. In this sense, we appeal to numerical schemes. Let us consider the following initial value problem:
\begin{eqnarray}
^{\rm C} D^\alpha_{t} y(t) &=& f(t, y(t)), \label{initialvalue1} \\
y^j (0) &=& y_0^{(j)} \label{initialvalue2}, 
\end{eqnarray}
where $t \in [0,T]$, with $T>0$, $j=0,1,..., n-1$, with $n = \lceil{\alpha}\rceil$ and $\alpha>0$. The fractional operator, in Caputo's sense, is:
\begin{equation}
^{\rm C} _0 D^\alpha_{t} y(t) = \frac{1}{\Gamma(n - \alpha)} \int_{0}^{t} (t-s)^{n-\alpha-1} \frac{d^n y}{ds^n} ds, \label{caputo}
\end{equation} 
with $n-1<\alpha<n$~\cite{Evangelista2023}. However, due to the nature of our problems, we restrict our analyses to $\alpha \in (0,1)$.

For the development of the numerical scheme, we consider the Riemann-Liouville fractional integral~\cite{Diethelm2010}, given by
\begin{equation}
J^\alpha_t g(t) = \frac{1}{\Gamma(\alpha)} \int_{0}^{t} (t-\omega)^{\alpha-1} g(\omega) d\omega. 
\label{riemann-integral}
\end{equation}
Once applied Eq. (\ref{riemann-integral}) in Eq. (\ref{caputo}), we obtain
\begin{equation}
J^\alpha_t D^{\alpha}_t y(t) = \frac{1}{\Gamma(\alpha)} \int_0^t (t -\omega)^{\alpha - 1} D^\alpha_\omega y(\omega) d\omega,
\end{equation}
substituting $D^\alpha_{t} y(t)$, we get
\begin{equation}
J^\alpha_t D^{\alpha}_t y(t) = \frac{1}{\Gamma(\alpha)\Gamma(1-\alpha)} \int_0^t dx \ y'(x) \left[\int_x^t (t-\omega)^{\alpha-1} (\omega-x)^{-\alpha} d\omega \right].
\end{equation}
At this point, we consider the transformation: $s = \omega - x$ and obtain
\begin{equation}
J^\alpha_t D^{\alpha}_t y(t) = \frac{1}{\Gamma(\alpha)\Gamma(1-\alpha)} \int_0^t dx \ y'(x) \left[\int_0^{t-x} s^{-\alpha} (t - x - s)^{\alpha-1} \ ds \right],
\end{equation}
which can be rewritten as
\begin{eqnarray}
 J^\alpha_t D^{\alpha}_t y(t) &=& \frac{1}{\Gamma(\alpha)\Gamma(1-\alpha)} \int_0^t dx \ y'(x) \nonumber \\ &\times& \left[\int_0^{t-x} s^{-\alpha} (t-x)^{\alpha-1} 
  \left(1 - \frac{s}{t -x} \right)^{\alpha-1} \ ds \right].
\end{eqnarray}
Now, we need to find a way to calculate the integral. One form is by introducing a new variable $\xi = s/(t - x)$, which results
\begin{equation}
J^\alpha_t D^{\alpha}_t y(t) = \frac{1}{\Gamma(\alpha)\Gamma(1-\alpha)} \int_0^t dx \ y'(x) \left[\int_0^{1} \xi^{-\alpha} (1-r)^{\alpha - 1} \ d\xi \right],
\label{resolver_beta}
\end{equation}
to solve the integral in $d\xi$, we consider the beta function, defined by 
\begin{equation}
\beta(p,q) = \int_0^1 dt \ t^{p-1} (1-t)^{q-1},
\label{beta}
\end{equation}
then
\begin{equation}
J^\alpha_t D^{\alpha}_t y(t) = \int_0^t y'(x) dx = y(t) - y(0).
\end{equation}
With this relation, we can write
\begin{equation}
y(t) = y(0) + J^\alpha_t D^{\alpha}_t y(t),
\end{equation}
obtaining
\begin{equation}
y(t) = y(0) + \frac{1}{\Gamma(\alpha)} \int_0^t (t-\omega)^{\alpha-1} g(\omega) \ d\omega.
\end{equation}
In discrete formulation,
\begin{equation}
y(t_{k+1}) = y(0) + \frac{1}{\Gamma(\alpha)} \int_0^{t_{k+1}} (t_{k+1} -\omega)^{\alpha-1} g(\omega) \ d\omega.
\label{aproximacao1}
\end{equation}
The solution gives the complete discrete form: 
\begin{eqnarray}
I_{k+1} &=& \int_0^{t_{k+1}} ds \ (t_{k+1} - s)^{\alpha - 1} \ \widetilde{g}_{k+1} (s), \\
&=& \sum_{j=0}^k \int_{t_j}^{t_{j+1}} ds \ (t_{k+1} - s)^{\alpha - 1} \ \widetilde{g}_{k+1} (s).
\end{eqnarray}
Solving the integral, we obtain:
\begin{equation}
I_{k+1} = \alpha^{-1} h^\alpha \sum_{j=0}^{k} f(t_j) [(k+1 - j)^{\alpha} - (k-j)^\alpha],
\end{equation}
substituting in Eq.~(\ref{aproximacao1}), the final form is
\begin{equation}
y_{k+1} = y_0 + \frac{h^\alpha}{\Gamma(\alpha + 1)} \sum_{j=0}^k [(k+1 - j)^{\alpha} - (k-j)^\alpha] f(t_j, y_j), \label{discreta}
\end{equation}
if $\alpha=1$, the Euler method for ODE (standard formulation) is recovered. Equation (\ref{discreta}) gives the numerical scheme for solving fractional differential equations. For our simulations, we consider $h=0.01$.

{
The convergence of the Fractional Euler method (Eq. (\ref{discreta})) is given by the following steps. Firstly, consider $y_j$ as solutions of Eq. (\ref{discreta}), where $j=1,2,...,k+1$. Assume that $f(t,y)$ (from Eq. (\ref{initialvalue1})) respects the Lipschitz condition in relation to $y$. Then, there is a Lipschitz constant $L$ in a given interval in which the solution is unique. If this is satisfied, the Eq. (\ref{discreta}) is stable. 

To prove the arguments, we include the perturbations $\tilde{y}_0$ in $y_0$ and  $\tilde{y}$ in $y_j$, then, by Eq. (\ref{discreta}) we have
\begin{equation}
    y_{k+1} + \tilde{y}_{k+1} = y_0 + \tilde{y}_0  + \sum_{j=0}^{k} b_{j,k} f (t_j, y_j + \tilde{y}_0) \label{perturbada},
\end{equation}
where $b_{j,k} = \left[(k+1-j)^\alpha - (k-j)^\alpha\right]h^\alpha / \Gamma(\alpha+1)$. Subtracting Eq. (\ref{perturbada}) from Eq. (\ref{discreta}) and taking the absolute value, we get
\begin{equation}
    | \tilde{y}_{k+1} | =  \bigg{|} \tilde{y}_0 + \sum_{j=0}^{k} b_{j,k} \left[f(t_j,y_j + \tilde{y}_j) - f(t_j,y_j) \right] \bigg{|}.
\end{equation}

The next step is to evaluate the component $j=0$ in the sum and rewrite:
\begin{equation}
  | \tilde{y}_{k+1} | =   \bigg{|} \tilde{y}_0 + b_{0,k} [f(t_0,y_0 + \tilde{y}_0) - f(t_0,y_0)] + \sum_{j=1}^{k} b_{j,k} \left[f(t_j,y_j + \tilde{y}_j) - f(t_j,y_j) \right] \bigg{|}.
\end{equation}
Using the Lipschitz condition and considering
\begin{equation}
    \eta_0 = {\rm max}_{0 \leq k \leq N} \left\{\sum_{j=0}^{n-1} \tilde{y}_0 + L b_{0,k} |\tilde{y}_0|  \right\},
\end{equation} 
we obtain
\begin{eqnarray}
    |\tilde{y}_{k+1}| &\leq& \eta_0 + \sum_{j=1}^{k} b_{j,k} \left[f(t_j,y_j + \tilde{y}_j) - f(t_j,y_j) \right], \\
    &\leq& \eta_0 + L \sum_{j=1}^k |\tilde{y}_j|,
\end{eqnarray}
using the Lemmas 3.1 and 3.4 from Ref. \cite{Li2013}, we get
\begin{equation}
    |\tilde{y}_{k+1}| \leq C \eta_0,
\end{equation}
where $C$ is a positive constant independent of $h$ and $k$.
}
%%%%%%%%%%%%%%%%%%%%%%%%%%%%%%%%%%%%%%
\subsection{Fractal operator}

In the diffusion context, some processes in complex media do not obey the Fick laws. In this sense, different model formulations were proposed~\cite{Liang2019}, such as diffusion equations based on Hausdorff derivative~\cite{Zhang2011}. Hausdorff derivative is known as fractal derivative and is not based on convolutions~\cite{Sun2013}. This derivative is defined in a non-Euclidean fractal metrics~\cite{Chen2017, Chen2006}, and its construction is not complicated once it is a local operator different from fractional ones.

We consider a one-dimensional problem of a given particle in movement to obtain a fractal derivative. If the time and space do not have fractal topology, the space and time interval are $\Delta x$ and $\Delta t$. However, if the movement is given in a fractal time, we have
\begin{equation}
\Delta t \rightarrow \Delta t^\alpha
\end{equation}
where $\alpha$ is the fractal time dimension. Considering that this particle start its movement in $t_0 = 0$ and moves with a velocity $v$, the displacement is 
\begin{equation}
s(t) = \int_0^t \ d\tau^\alpha \ v(\tau).
\end{equation}
Differentiating this expression, we obtain $ds/dt^\alpha$, 
{

which can be written, in terms of the limit condition, as follows
\begin{equation}
\frac{ds}{dt^\alpha} = \lim_{t_1 \rightarrow t} \frac{s(t_1) - s(t)}{(t_1 - t_0)^\alpha - (t - t_0)^\alpha},
\end{equation}
choosing $t_0 = 0$, we get
\begin{equation}
\frac{ds}{dt^\alpha} = \lim_{t_1 \rightarrow t} \frac{s(t_1) - s(t)}{t_1^\alpha - t^\alpha}.
\label{fractal_definition0}
\end{equation}

Equation (\ref{fractal_definition0}) defines the Hausdorff derivative. The connection between fractal and standard derivative can be obtained by the chain rule, as follows:
\begin{equation}
    \frac{ds}{dt^\alpha} = \frac{ds}{dt} \frac{dt}{dt^\alpha} = \frac{1}{\alpha} t^{1 - \alpha} \frac{ds}{dt}.
    \label{fractal_definition}
\end{equation}
To evaluate $dt / dt^\alpha$, we consider the inverse function theorem.

Suppose the following initial value problem
\begin{eqnarray}
    \frac{d s}{dt^\alpha} &=& f(t,s(t)), \label{initial_value1} \\ 
    s(0) &=& s_0, 
\end{eqnarray}
one way to solve this problem is to use Eq. (\ref{fractal_definition}) and rewrite the initial value problem in
\begin{eqnarray}
    \frac{ds}{dt} &=& \alpha t^{\alpha-1} f(t,s(t)), \\ 
    s(0) &=& s_0, 
\end{eqnarray}
which can be solved by standard methods. In this work, we solve using the Runge-Kutta 4th method.
}
%%%%%%%%%%%%%%%%%%%%%%%%%%%%%%%%%%%%%%%%%%%%%%%%
\section{Epidemiological models}
Compartmental models compartmentalize the host population ($N$) according to the stage of infection~\cite{AndersonMay}. The main compartments are susceptible ($S$), infected ($I$), exposed ($E$) and recovered ($R$)~\cite{Bjornstad2018}. The $S$ compartment is responsible for holding the individuals who can be infected. The $E$ compartment keeps the individuals who were infected and did not transmit the infection (latent period)~\cite{Sharma2021} or transmit with a lower probability when compared with $I$ individuals (incubation period)~\cite{Amaku2021b}. The $I$ compartment contains the individuals who transmit the infection, i.e., the infectious individuals. Finally, the $R$ compartment collects the individuals who passed the infectious period. The epidemics models are obtained by compositions of these compartments, which depend on the disease properties. The main models are SI, SIS, SIR, and SEIR. In the next sections, we discuss each one in more detail and its applications.
%%%%%%%%%%%%%%%%%%%%%%%%%%%%%%%%%%%%%%%%%%%%%%%%%%%
\subsection{Susceptible--Infected (SI) model}
%%%%%%%%%%%%%%%%%%%%%%%%%%%%%%%%%%%%%%%%%%%%%%%%%%%
\subsubsection{SI standard model}\label{si_standard}

The SI model describes the interaction between $S$ and $I$ individuals in a given population with size $N$, where $S+I=N$ \cite{batista2021_RBEF}. The population is supposed to be homogeneous and well-mixed. The transmission depends on the contacts between $I$ and $S$ individuals as well as the characteristics of the illness. The transmission is namely contacts, and we are supposed to be constant. The multiplication of contacts by $S/N$ gives us the rate at which the susceptible individuals are contacted by the infected, which is
\begin{equation}
	\frac{contacted \  individual}{infected \ individual \cdot time} \cdot \frac{S(t)}{N} = \frac{susceptible \ contacted}{infected \ individual \cdot time}.
	\label{rel1}
\end{equation}
Multiplying this relation by the probability of each susceptible acquiring the disease due to the contact, we obtain a new relationship that gives us the rate of newly infected per infected individual:
\begin{eqnarray}
	&&\frac{new \ infected}{infected \ individual \cdot time} = \nonumber \\
	&&\frac{contacted \  individual}{infected \ individual \cdot time} \cdot \frac{S(t)}{N} \cdot {transmission \ probability},
	\label{rel2}
\end{eqnarray}
from which it is possible to define the constant $\beta$, that is
\begin{equation}
	\beta \equiv  \frac{contacted \  individual}{infected \ individual \cdot time} \cdot {transmission \ probability}.
	\label{rel3}
\end{equation}
This constant is named {\it coefficient transmission} and is the rate of new infections when all contacted individuals are susceptible. Note that $\beta$ is the product of two quantities. The first is associated with the behavior of the individuals, and the second is a characteristic of the illness. 

\begin{figure}[!hbt]
	\centering 
	\includegraphics[scale=0.5]{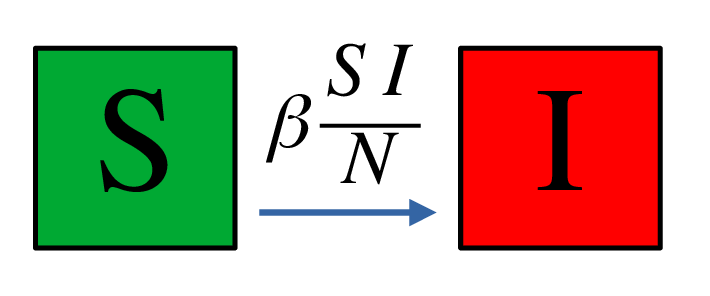} 
	\caption{Schematic representation of SI model.}
	\label{fig1}
\end{figure}

Considering $\beta$ defined by Eq. (\ref{rel3}), it is possible to rewrite Eq. (\ref{rel2}) in a compact form. However, before that, we multiply Eq. (\ref{rel2}) by the infected number at $t$ time ($I(t)$), obtaining 
\begin{equation}
	\frac{new \ infected}{time} = \beta \cdot \frac{S(t)}{N} \cdot I(t) \equiv \frac{dI}{dt}.
 \label{rateI}
\end{equation}
This relation shows us how the infected rate (${dI}/{dt}$) increases. As the total population is $N = S + I$, as $I$ increases, $S$ decreases. Then, a flux transition occurs between the compartments, represented in Fig. \ref{fig1}. The constant $N$ is a constraint in the system.

\begin{figure}[!hbt]
	\centering 
	\includegraphics[scale=1.0]{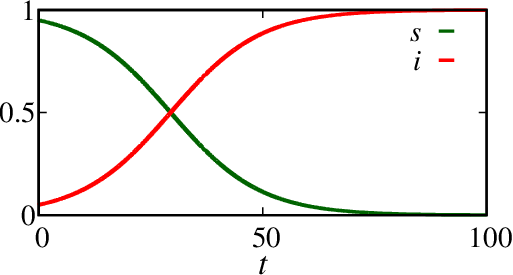} 
	\caption{Time evolution of SI standard model. The green and red lines 
	show the fraction of $s$ and $i$ individuals, respectively. We consider
	$\beta=0.1$, and $i_0 = 0.05$.}
	\label{fig2}
\end{figure}

To remain $N=const.$, the quantity that increases in $I$, decreases from $S$, then the SI model is described by the following ordinary differential equations (ODE)
\begin{eqnarray}
	\frac{dS}{dt} &= -\beta\frac{SI}{N} \label{si-eq1}, \\
	\frac{dI}{dt} &= \beta\frac{SI}{N}. \label{si-eq2}
	\label{si-model}
\end{eqnarray}
These equations can be written in a normalized way by replacing $S \rightarrow Ns$ and $I \rightarrow Ni$, which results in
\begin{eqnarray}
\frac{ds}{dt} &=& -\beta{si}, \label{si-stdn-eq1} \\
\frac{di}{dt} &=& \beta{si}, \label{si-stdn-eq2}.
\end{eqnarray}
{ Eqs. (\ref{si-stdn-eq1}) and (\ref{si-stdn-eq2}) together with the initial conditions $s(0) = s_0$ and $i(0) = i_0$ defined an initial value problem. For biological motivations, we are only interested in solutions inside $D = \{(s,i) \in [0,1]^2 \ |  \ s(t) + i(t) = 1 \}$. Once, $s_0 \geq 0$, $i_0 \geq 0$, and $\beta \geq 0$, the solutions are positive defined for all $t \geq 0$ \cite{Gao2008, Gabrick2023c}.
}
The equilibrium solutions are $(\widehat{s},\widehat{i}) = (1,0)$ and $(s^{*},i^{*})=(0,1)$. The first one is called disease-free, and the 
second is the endemic solution. Considering the constrain, Eqs.~(\ref{si-stdn-eq1}) and (\ref{si-stdn-eq2}) are replaced just by one equation, which is
\begin{equation}
	\frac{di}{dt} = \beta (1-i)i,
\end{equation}
with solution
\begin{equation}
	i(t) = \frac{i_{0} e^{\beta t}}{1 + i_{0}(e^{\beta t} - 1)}, \label{sol-si}
\end{equation}
and $s = 1 - i$. If $\beta>0$ then $\lim_{t \rightarrow \infty} i(t) = 1$ and $\beta = 0$, then $i(t) = i_{0}$. In this way, if $\beta>0$, the non-trivial equilibrium is found when all population is infected. Considering $\beta=0.1$, a time series is displayed in Fig.~\ref{fig2}. The green and red lines are related to $s$ and $i$, respectively. In this result, we consider $i_{0}=0.05$. As time advances, the fraction of individuals related to $s$ starts to move to the $i$ compartment. Then, as one is emptied, the other is filled.

%%%%%%%%%%%%%%%%%%%%%%%%%%%%%%%%%%%%%%%%%%%%%%%%%%%
\subsubsection{SI fractional model} \label{fractional_si}

\begin{figure}[!hbt]
	\centering
	\includegraphics[scale=1.0]{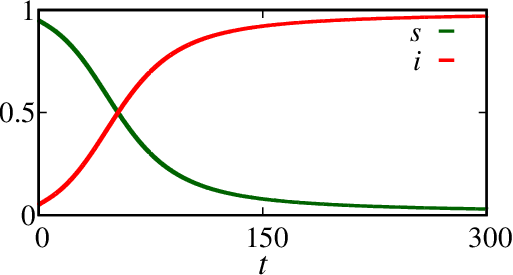} 
	\caption{Time evolution of fractional SI model. The green and red lines 
	show the fraction of $s$ and $i$ individuals for $\alpha=0.8$, respectively. We consider 
	$\beta=0.1$, and $i_0 = 0.05$.}
	\label{fig3}
\end{figure}

Before introducing the fractional extension of the SI model, we analyze the dimensions of the equations. The unities are: $t \ = \ [time]$, $S, I, N \ = \ [individual]$, $\beta \ = \ [time]^{-1}$. Then, Eqs.~(\ref{si-eq1}) and (\ref{si-eq2}) have unity of $ind. / time$. To respect the unities of the model, when we consider fractional operators, it is necessary to include a correction in the constant, that is $\beta^\alpha$, resulting in $ind./ time^\alpha$~\cite{Chen2021, Diethelm2013}. In this way, the normalized Eqs.~(\ref{si-stdn-eq1}) and (\ref{si-stdn-eq2}) are written in fractional form as follows
\begin{eqnarray}
\frac{d^\alpha s}{dt^\alpha} &=& -\beta^\alpha {si}, \label{si-frac-eq1} \\
\frac{d^\alpha i}{dt^\alpha} &=&  \beta^\alpha  {si}. \label{si-frac-eq2}
\end{eqnarray}
In this Chapter, to verify the differences between the formulations, we consider the same parameter value, i.e., $\beta = \beta^\alpha$. Then, to economy notation, we refer to $\beta$, except in this Section. 
{ Considering Caputo's definition, the solutions for Eqs. (\ref{si-frac-eq1}) and (\ref{si-frac-eq2}) are restrict to $D = \{(s,i) \in [0,1]^2 \ |  \ s(t) + i(t) = 1 \}$, once $s_0$, $i_0$, $\alpha$, and $\beta$ are greater than or equal to zero for $t\geq 0$. The complete proof can be found in Ref. \cite{Balzotti2021}.}
In Caputo's definition, the derivative of a constant remains equal to zero. In this way, 
the constraint is respected ($s+i=1$), and the equilibrium is the same as in the standard case. Besides that, a solution type Eq.~(\ref{sol-si}) is not possible due to the characteristics of fractional operators. Numerical solutions for these equations can be obtained by the method described in Section~\ref{Fractional_scheme}. The result for $\alpha=0.8$ and $\beta = 0.1$ is displayed in Fig.~\ref{fig3}. In this formulation, the global aspect of the solutions remains unchanged. However, the disease takes a long time to reach the equilibrium. In the standard case, the equilibrium is found in $\tau=75.4$ and in the fractional approach $\tau=895.97$. For numerical reasons, we consider that the equilibrium is found when $i(t)=0.99$. 

\begin{figure}[!hbt]
	\centering
	\includegraphics[scale=1.0]{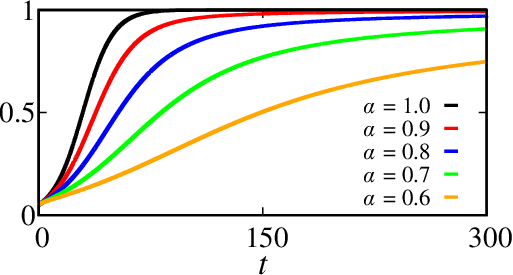} 
	\caption{Comparison between standard model (black line) and fractional 
	model for $\alpha=0.9$ (red line), $\alpha=0.8$ (blue line), $\alpha=0.7$ 
	(green line) and $\alpha=0.6$ (orange line). 
	We consider $\beta=0.1$, and $i_0 = 0.05$.}
	\label{fig4}
\end{figure}

Figure~\ref{fig4} displays different infected curves for $\alpha=1$ (standard model) by the black line, $\alpha=0.9$ by the red line, $\alpha=0.8$ by the blue line, $\alpha=0.7$ by the green line and $\alpha=0.6$ by the orange line. These results show that how less is $\alpha$ less is the velocity at which the infected curve increases. Consequently, more time is spent to reach the equilibrium point. 

%%%%%%%%%%%%%%%%%%%%%%%%%%%%%%%%%%%%%%%%%%%%%%%%%%%
\subsubsection{SI fractal model}

\begin{figure}[!hbt]
	\centering
	\includegraphics[scale=1.0]{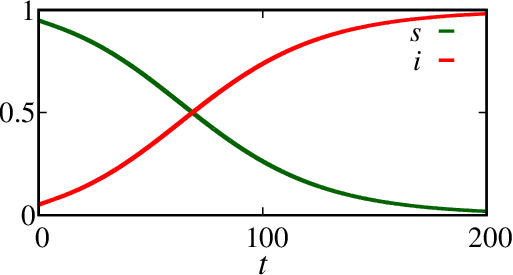} 
	\caption{Time evolution of fractal SI model for $\alpha = 0.8$ 
	We consider $\beta=0.1$, and $i_0 = 0.05$.}
	\label{fig5}
\end{figure}

To construct the fractal formulation of the SI model, we follow the same strategy considered in Section~\ref{fractional_si}. To respect the dimensions of the 
system, the unit of $\beta$ needs to be reformulated to $\beta$. Considering Eq.~(\ref{fractal_definition}), Eqs.~(\ref{si-stdn-eq1}) and~(\ref{si-stdn-eq2}) become
\begin{eqnarray}
	\frac{ds}{dt^{\alpha}} \equiv \frac{ds}{dt} &=& - \alpha t^{\alpha-1} \beta si, \label{fractal_si-1} \\
	\frac{di}{dt^{\alpha}} \equiv \frac{di}{dt} &=& \alpha t^{\alpha-1} \beta si. \label{fractal_si-2}
\end{eqnarray}
{ We consider solutions restrict into $D = \{(s,i) \in [0,1]^2 \ |  \ s(t) + i(t) = 1 \}$, with $s_0 \geq 0$, $i_0 \geq 0$, $\alpha\geq0$, $t \geq0$, and $\beta\geq0$.
}
Summing the equations, we obtain $s+i=1$. Due to the direct connection with standard derivatives a closed solution form is possible. The solution is
 \begin{equation}
	i(t) = \frac{i_{0} e^{\beta t^{\alpha}}}{1 + i_{0}(e^{\beta t^{\alpha}} - 1)}, \label{sol-si-frac}
\end{equation}
and $s(t) = 1 - i(t)$. { As the parameters are positive, the Eq. (\ref{sol-si-frac}) and $s(t) = 1 - i(t)$ are always positive}. The properties of the fractal calculus with the standard model are very similar; in fact, it comes from the direct relation by integer derivatives. A time series for Eq.~(\ref{sol-si-frac}) is shown in Fig.~\ref{fig5} for $\alpha=0.8$. In this formulation, the epidemic curve increases slower than in the standard case and faster than fractional. Due to the nature of time, i.e., $t^{\alpha}$, the initial behavior of spread, which is dominated by the exponential, increases with less velocity as measures $\alpha$ decrease. In this case, the equilibrium is reached in $\tau=222.25$.

\begin{figure}[!hbt]
	\centering
	\includegraphics[scale=1.0]{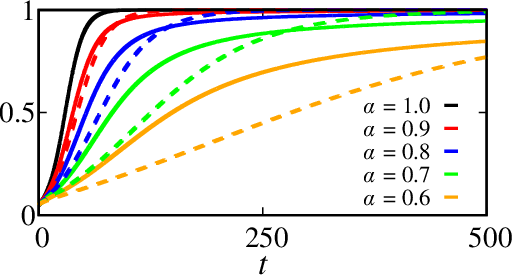} 
	\caption{Comparison between the three formulations for the SI model: 
	black line standard model, continuous fractional, and dotted fractal approach.  
	We consider $\beta=0.1$, and $i_0 = 0.05$.}
	\label{fig6}
\end{figure}

A comparison between the three formulations is displayed in Fig.~\ref{fig6}. The black line is for the standard case, while the colorful ones are for the fractional (continuous) and fractal (dotted). The red lines are for $\alpha=0.9$, the blue lines for $\alpha=0.8$, the green lines for $\alpha=0.7$, and the orange lines for $\alpha=0.6$. Now the difference between the time evolution of the SI model governed by the three formulations is clear. The fractional always evolves with a lower velocity; conversely, the fractal formulation evolves with a lower velocity for short times. The equilibrium is reached earlier. The implications of this 
different approach are mainly in the transient time ($\tau$) and how it is reached. 

\begin{figure}[!hbt]
	\centering
	\includegraphics[scale=1.0]{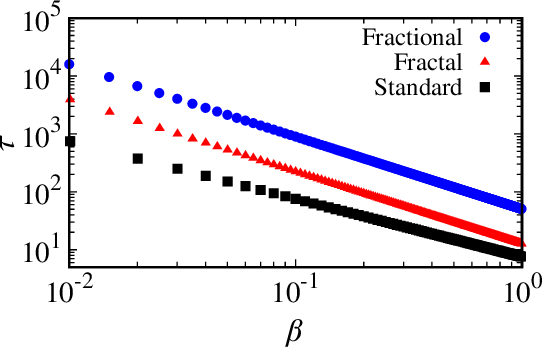} 
	\caption{Time to reach the steady solution ($\tau$) as a function of $\beta$ 
	for different formulations with $\alpha=0.8$: fractional (circles), 
	fractal (triangles), and standard (square).}
	\label{fig7}
\end{figure}

One way to measure the transient time is to consider a steady solution when $i(t) =0.99$. For the initial condition $i_0 = 0.05$ and $\alpha=0.8$, a numerical result is 
exhibited in Fig.~\ref{fig7}. The blue circles are associated with the fractional model, while the red triangles and black squares are related to the fractal and standard formulations. Depending on the formulation, the time to reach the equilibrium varies by one order of magnitude.

%%%%%%%%%%%%%%%%%%%%%%%%%%%%%%%%%%%%%%%%%%%%%%%%%%%
\subsubsection{Application of SI model}

\begin{figure}[!hbt]
	\centering
	\includegraphics[scale=1.0]{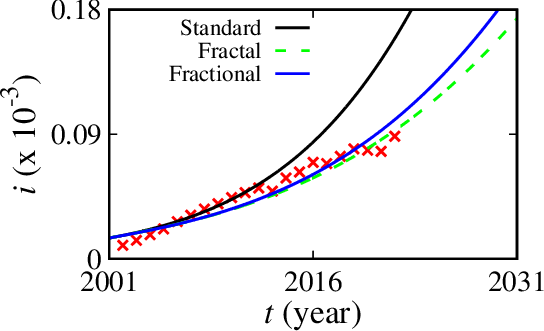} 
	\caption{Description of the AIDS case in three formulations: standard (black 
	line), fractal (dotted green line), and fractional (blue line). We 
	consider $i_0 = 0.015$ and $\beta = 0.12$.}
	\label{fig8}
\end{figure}

One application of the SI model is to study AIDS case evolution~\cite{Huang2005, Ghosh2018}. In this Chapter, as an application of SI model, we consider AIDS data from Bangladesh from 2001 to 2021\footnote{Available on: https://aphub.unaids.org/ Accessed on: 07 Jul. 2023.}. To use the models described by Eqs. (\ref{sol-si}) (standard), 
(\ref{si-frac-eq1})--(\ref{si-frac-eq2}) (fractional) and (\ref{sol-si-frac}) (fractal), we use fractions of infected individuals about the population size each year. For the standard model, our adjustment suggests $i_0 = 0.015$ and $\beta = 43.8$ (1/year), as shown through the black line in Fig.~\ref{fig8}. Fixing these parameters, an improvement is found in the data description for $\alpha=0.91$ in the fractal (green dotted line) and $\alpha=0.9$ in the fractional formulation (blue line). { In this work, we first consider the standard model and obtain parameters to describe the real data better. In cases where analytical solutions are possible, we just fit using the software Gnuplot 5.4.5. When it is not possible to obtain an analytical solution, we vary the parameters until obtain a better description. After getting the parameters, we change the order and the framework, to verify improvement or not.} In this way, the results present in Fig. \ref{fig8} suggest that the fractal formulation is better for describing the considered data.

%%%%%%%%%%%%%%%%%%%%%%%%%%%%%%%%%%%%%%%%%%%%%%%%%%%
%%%%%%%%%%%%%%%%%%%%%%%%%%%%%%%%%%%%%%%%%%%%%%%%%%%
\newpage
%%%%%%%%%%%%%%%%%%%%%%%%%%%%%%%%%%%%%%%%%%%%%%%%%%%
\subsection{Susceptible--Infected--Susceptible (SIS) model}
%%%%%%%%%%%%%%%%%%%%%%%%%%%%%%%%%%%%%%%%%%%%%%%%%%%
\subsubsection{SIS standard model}

\begin{figure}[!hbt]
	\centering
	\includegraphics[scale=0.5]{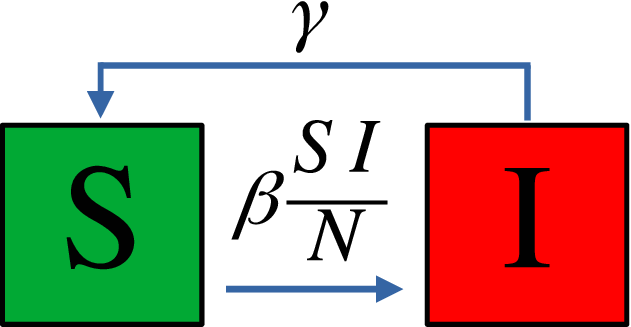}
	\caption{Schematic representation of SIS model.}
	\label{fig9}
\end{figure}

Some diseases do not confer long-life infection, such as gonorrhea~\cite{Hethcote1984} or syphilis~\cite{Gabrick2023c}. To model these diseases is necessary to adapt the SI model by including a new parameter, $\gamma$. The parameter $\gamma$ takes infected individuals and sends them to the susceptible compartment. This modification, the SIS model, is schematically represented in Fig.~\ref{fig9}.

The ODE formulation of the SIS model is very similar to the SI model (Eqs.~(\ref{si-eq1}) and (\ref{si-eq2})). The difference is removing a fraction $\gamma$ of the $I$ compartment and adding the same rate in the $S$ compartment. In this case, the equations become
\begin{eqnarray}
\frac{dS}{dt} &=& -\beta\frac{SI}{N} + \gamma I, \label{sis-model-s}\\
\frac{dI}{dt} &=& \beta\frac{SI}{N} - \gamma I. \label{sis-model-i}
\end{eqnarray}
If $t$ is measured in days, $dS/dt$ has individual/day unit, as well as $N$, $S$ and $I$ have unity of individual, then $\beta$ and $\gamma$ are 1/day unity. 

The $\gamma$ interpretation can be obtained by considering $S(t) =0$ and $I(0) =N$; with this consideration and after some calculus, we show that the average time for the individual reaches a healthy state again is $1/\gamma$ namely the average infectious period. Therefore, $\gamma$ is the recovery rate~\cite{Keeling2008}. 

Considering a normalized system, Eqs.~(\ref{sis-model-s}) and (\ref{sis-model-i}) becomes
\begin{eqnarray}
\frac{ds}{dt} &=& -\beta si + \gamma i, \label{sis-model-s-norm}\\
\frac{di}{dt} &=& \beta si - \gamma i. \label{sis-model-i-norm}
\end{eqnarray}
{ Given the initial conditions $s(0)=s_0$ and $i(0)=i_0$ greater than or equal to zero, the considered solutions for Eqs. (\ref{sis-model-s}) and (\ref{sis-model-i}) are defined in $D = \{(s,i) \in [0,1]^2 \ |  \ s(t) + i(t) = 1 \}$ with $\beta$ and $\gamma \geq 0$.} 
The equilibrium solutions for this system are: $(\widehat{s},\widehat{i}) = (1,0)$,  $(s^{*},i^{*})=(0,1)$ and $(\widetilde{s},\widetilde{i}) = \left(\gamma/\beta, 1 - \gamma/\beta\right)$. Equation (\ref{sis-model-i-norm}) can be rewritten as
\begin{equation}
\frac{di}{dt} = i \beta \left(s - \frac{\gamma}{\beta}\right).
\end{equation}
If $s_0 < \gamma/\beta$, the disease die out, once $di/dt<0$. On the other hand, if $s_0 > \gamma/\beta$, then $di/dt>0$ and the disease invades the population. This result is very important in mathematical epidemiology and is known as the ``threshold phenomenon"~\cite{Keeling2008}. The inverse of this rate is called  basic reproductive ratio ($R_0 \equiv {\beta}/{\gamma}$). If we assume $s_0$ near 1, the pathogen invades the population only if $R_0 > 1$.

\begin{figure}[!hbt]
	\centering 
	\includegraphics[scale=1]{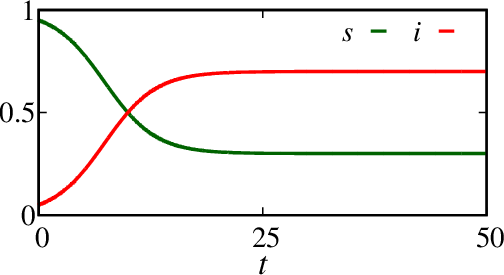}
	\caption{Time series for $s$ (green line) and $i$ (red line). We 
	consider $\beta = 0.5$, $\gamma=0.15$, and $i_0 = 0.05$.}
	\label{fig10}
\end{figure}

Considering the constraints, we can write 
\begin{equation}
\frac{di}{dt} = i \beta (1 - i - R_0^{-1}),
\end{equation}
with the solution given by 
\begin{equation}
i(t) = \frac{i_0 \xi e^{\xi \beta t}}{\xi + i_0 (e^{\xi \beta  t} - 1)},
\label{sis-analitica}
\end{equation}
where $\xi = 1 - R_0^{-1}$ and $s(t) = 1 - i(t)$. If $R_0 < 1$, then $\xi < 0$ and the solution goes to $(\widehat{s},\widehat{i})$. However, if $R_0>1$ then $\xi>0$, and the solution goes to $1 - R_0^{-1}$ in the limit of $t \rightarrow \infty$. Figure \ref{fig10} displays the time series for $s$ in the green line and $i$ in the red line. The initial behavior is similar to the SI model. The $s$ curve decays as the $i$ curve increases. However, the infected curve reaches a stationary evolution after a certain time. In this case, the individuals in the $s$ compartment are transferred to the $i$ compartment at the same rate at which the individuals lost the infection. Due to this fact, both lines are parallel.  

\begin{figure}[!hbt]
	\centering 
	\includegraphics[scale=0.55]{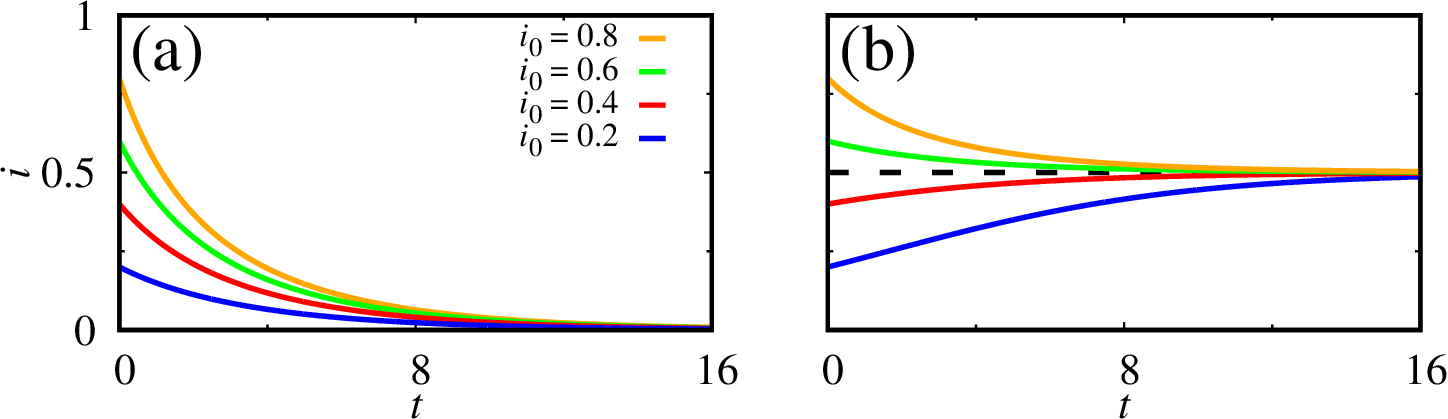}
	\caption{Solutions for SIS model with different initial conditions.
	The parameters in panel (a) are $\beta = 0.25$, $\gamma = 0.5$, and $R_0 = 0.5$.
	In the panel (b) the parameters are $\beta=0.5$, $\gamma=0.25$ and $R_0 =2$.}
	\label{fig11}
\end{figure}

Figure~\ref{fig11} displays the time series of $i$ variable for $i_0 = 0.8$ (orange line), $i_0 = 0.6$ (green line), $i_0 = 0.4$ (red line) and $i_0 = 0.2$ (blue line). The panel (a) is for $R_0 = 0.5$ with $\beta = 0.25$ and $\gamma = 0.5$. The panel (b) is for $R_0 = 2$ with $\beta = 0.5$ and $\gamma = 0.25$. The solutions in panel (a) show the infected number going to zero for different initial conditions, as suggested by our analyses when $R_0 <1$. On the other hand, panel (b) exhibits the solutions when $R_0 > 1$; in this case, the asymptotic solution is $1 - R_0^{-1}$ (black line).

%%%%%%%%%%%%%%%%%%%%%%%%%%%%%%%%%%%%%%%%%%%%%%%%%%%
\subsubsection{SIS fractional model} \label{section_fractional_sis}

\begin{figure}[!hbt]
	\centering 
	\includegraphics[scale=1]{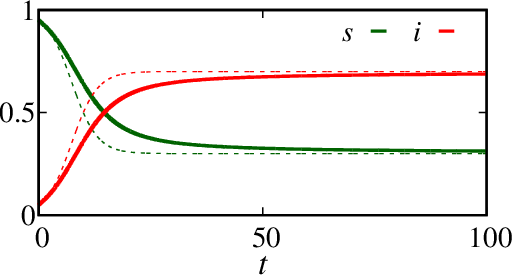}
	\caption{Time series for $s$ (green line) and $i$ (red line) with $\alpha=0.8$. 
	The dotted green and red lines show the $s$ and $i$, respectively, for the 
	standard case. We consider $\beta = 0.5$, $\gamma=0.15$, and $i_0 = 0.05$.}
	\label{fig12}
\end{figure}

The fractional extension of the SIS model is obtained by changing the operators in Eqs.~(\ref{sis-model-s-norm}) and (\ref{sis-model-i-norm}), which leads to 
\begin{eqnarray}
\frac{d^\alpha s}{dt^\alpha} &=& -\beta si + \gamma i, \label{fractional_sis-s}\\
\frac{d^\alpha i}{dt^\alpha} &=& \beta si - \gamma i. \label{fractional_sis-i}
\end{eqnarray}
{ The solutions for Eqs. (\ref{fractional_sis-s}) and (\ref{fractional_sis-i}) are defined into $D = \{(s,i) \in [0,1]^2 \ |  \ s(t) + i(t) = 1 \}$ with $s_0$, $i_0$, $\alpha$, $\beta$, and $\gamma\geq0$.} Due to the adjustment in the unities, the basic reproductive number is $R_0 = \gamma^\alpha/\beta^\alpha$, or $R_0 = \gamma/\beta$. The modification in the equilibrium solution is only in the correction of the unities. Due to the fractional operators' nature, it is impossible to obtain a close solution as in the standard case. Figure~\ref{fig12} displays a time series for $s$ (green line) and $i$ (red line) for $\beta=0.5$, $\gamma=0.15$, $\alpha=0.8$ and $i_0 = 0.05$. The dotted lines indicate the solutions for $\alpha=1$. The steady state is the same for fractional and integer cases since we choose the same value of parameters. The individuals transferring from one compartment to another is slower when compared with the standard case, and the steady solution takes more time to be reached.

\begin{figure}[!hbt]
	\centering
	\includegraphics[scale=1.0]{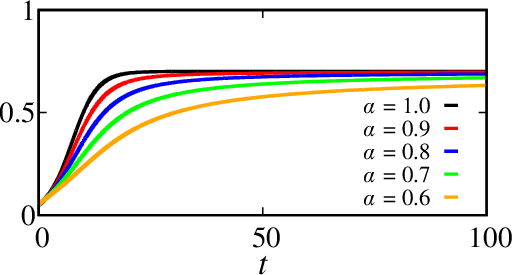} 
	\caption{Comparison between standard (black line) and fractional SIS 
	model for $\alpha=0.9$ (red line), $\alpha=0.8$ (blue line), $\alpha=0.7$ 
	(green line) and $\alpha=0.6$ (orange line). 
	We consider $\beta=0.5$, $\gamma=0.15$, and $i_0 = 0.05$.}
	\label{fig13}
\end{figure}

For fixed $\beta$ and $\gamma$, Fig.~\ref{fig13} shows the different behaviors of $i$ for $\alpha=1$ (black line), $\alpha=0.9$ (red line), $\alpha=0.8$ (blue line), $\alpha=0.7$ (green line) and $\alpha = 0.6$ (orange line). These results show that the initial behavior, which is exponential in the standard case, increases slowly when 
$\alpha$ is decreased. In the limit $t \rightarrow \infty$, the solutions converge to the steady state defined by $(\widetilde{s},\widetilde{i})$.

%%%%%%%%%%%%%%%%%%%%%%%%%%%%%%%%%%%%%%%%%%%%%%%%%%%
\subsubsection{SIS fractal model}

\begin{figure}[!hbt]
	\centering 
	\includegraphics[scale=1]{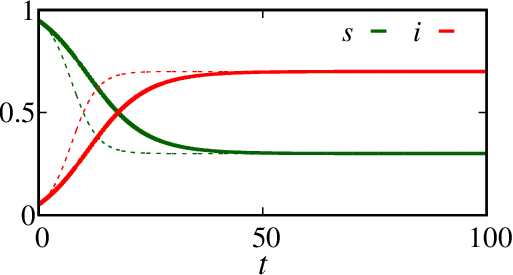}
	\caption{Time series for SIS model. The green and red lines are related 
	to $s$ and $i$ with $\alpha=0.8$. 
	The dotted green and red lines show the $s$ and $i$, respectively, for the 
	standard case. We consider $\beta = 0.5$, $\gamma=0.15$ and $i_0 = 0.05$.}
	\label{fig14}
\end{figure}

A fractal extension of the SIS model is given by 
\begin{eqnarray}
	\frac{ds}{dt} &=& \alpha t^{\alpha-1} \left(-\beta si + \gamma i\right), \label{fractal_sis-1} \\
	\frac{di}{dt} &=& \alpha t^{\alpha-1} \left(\beta si - \gamma i \right), \label{fractal_sis-2}
\end{eqnarray}
{ where the solutions are restricted to $D = \{(s,i) \in [0,1]^2 \ |  \ s(t) + i(t) = 1 \}$ and the parameters $s_0$, $i_0$, $\alpha$, $\beta$, and $\gamma$ are positive.}
The basic reproduction rate, in this case, is
\begin{equation}
R_0 = \frac{\beta}{\gamma}.
\end{equation}
Considering the constrain $s+i=1$, it is possible to write
\begin{equation}
\frac{di}{dt} = \alpha t^{\alpha-1} i \beta \left(1 - i - R^{-1}_0 \right),
\end{equation}
where the solution is
\begin{equation}
i(t) = \frac{i_0 \xi e^{\xi \beta t^\alpha }}{\xi + i_0 (e^{\xi \beta t^\alpha} - 1)},
\end{equation}
with $\xi = 1 - R^{-1}_0$ and $s(t) = 1 - i(t)$. A time series is exhibited in Fig.~\ref{fig14} for $\alpha=0.8$. The green and red lines are related to $s$ and $i$, respectively. The dotted line corresponds to the standard model. The difference about the standard model is in the transient time, i.e., the time before reaching the steady solution $(\widetilde{s},\widetilde{i})$. For $\alpha<1$, the transference between the compartments is decreased as $\alpha$ function. 

\begin{figure}[!hbt]
	\centering
	\includegraphics[scale=1.0]{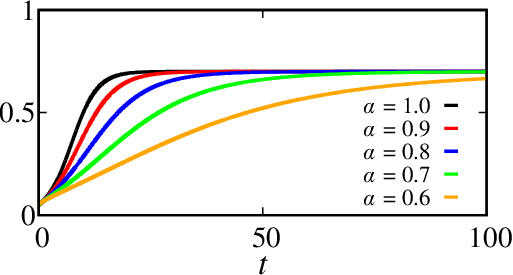} 
	\caption{Comparison between standard (black line) and fractal SIS 
	model for $\alpha=0.9$ (red line), $\alpha=0.8$ (blue line), $\alpha=0.7$ 
	(green line) and $\alpha=0.6$ (orange line). 
	We consider $\beta=0.5$,  $\gamma=0.15$ and $i_0 = 0.05$.}
	\label{fig15}
\end{figure}

Figure~\ref{fig15} displays the $i$ solutions for different $\alpha$ values. The black line is for the standard case, the red line for $\alpha=0.9$, the blue line for $\alpha=0.8$, the green line for $\alpha=0.7$, and the orange line for $\alpha=0.6$. As $\alpha$ decreases, the infected line increases with less intensity. Due to this fact, reaching a steady solution takes longer. 

\begin{figure}[!hbt]
	\centering
	\includegraphics[scale=1.0]{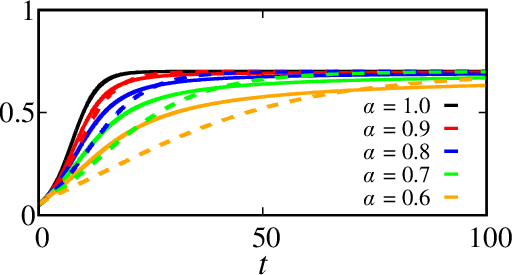} 
	\caption{Comparison between the three formulations for the SIS model: 
	black line standard model, continuous fractional, and dotted fractal approach.  
	In value, the constants are $\beta=0.5$, $\gamma=0.15$, and $i_0 = 0.05$.}
	\label{fig16}
\end{figure}

A comparison among the three formulations is shown in Fig.~\ref{fig16} for the same parameter values. The continuous black line is for the standard model, the continuous colorful lines are for the fractional approach, and the dotted lines are for the fractal approach. The difference is clear before the transient time. The fractal formulation increases with a lower velocity than the other two cases; however, it reaches the steady solution earlier than the fractional approach. 

%%%%%%%%%%%%%%%%%%%%%%%%%%%%%%%%%%%%%%%%%%%%%%%%%%%
\subsubsection{Application of SIS model}

\begin{figure}[!hbt]
	\centering
	\includegraphics[scale=1.0]{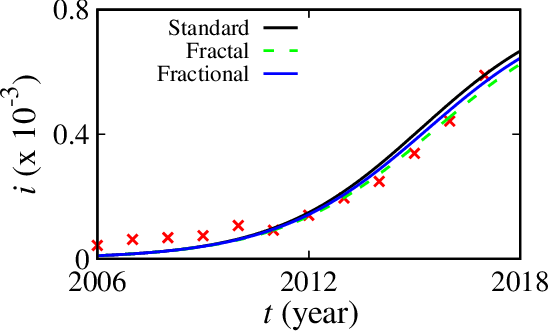} 
	\caption{Description of Syphilis cases in three formulations: SIS standard (black 
	line), SIS fractal (dotted green line), and SIS fractional (blue line). We 
	consider $\beta=208.05$, $\gamma=32.85$ (1/year), $i_0 = 0.01$ and $R_0 = 6.3$.}
	\label{fig17}
\end{figure}

As an application of the SIS model, we consider the syphilis data from Brazil (2006-2017). The data is available on ``Indicadores de Inconsist\^encias de S\'ifilis nos Munic\'ipios Brasileiros"\footnote{URL: http://indicadoressifilis.aids.gov.br/ Accessed on: 28 Jun. 2023.}. Moreover, we take the fraction of infected about the total population per year. Considering Eq.~(\ref{sis-analitica}) to estimate the parameters, we obtain $\beta=208.05$, $\gamma=32.85$ (1/year), $i_0 = 0.01$ and $R_0 = 6.3$. The results are displayed in Fig.~\ref{fig17}, where the red points are the data, the black line is the result obtained for the standard case, the dotted green line for the fractal approach ($0.98$), and the blue line for the fractional model ($0.98$). { Note that the three formations describe the data with good accuracy. Until 2012, the three approaches gave practically the same answer. However, after 2012, the fractal formulation is more close to the experimental points than the standard or fractional models.}

%%%%%%%%%%%%%%%%%%%%%%%%%%%%%%%%%%%%%%%%%%%%%%%%%%%
\subsection{Susceptible-Infected-Recovered (SIR) model}
%%%%%%%%%%%%%%%%%%%%%%%%%%%%%%%%%%%%%%%%%%
\subsubsection{SIR standard model}

\begin{figure}[!hbt]
	\centering
	\includegraphics[scale=0.4]{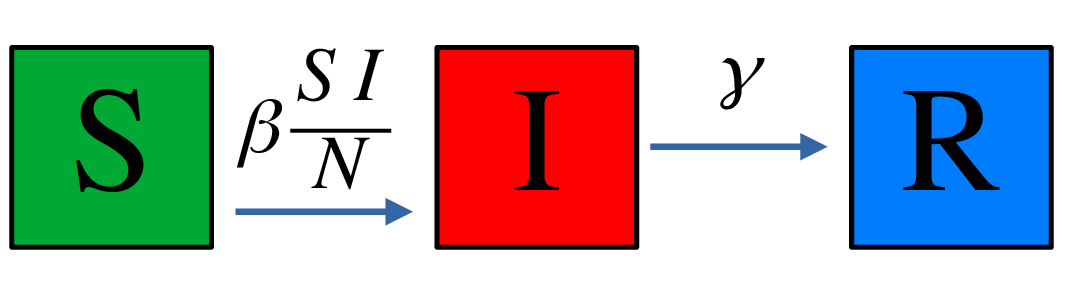} 
	\caption{Schematic representation of SIR model.}
	\label{fig18}
\end{figure}

The class of models discussed above describes the spread of disease that does not confer immunity to the individuals~\cite{Keeling2008}. However, it is known that some illnesses confer immunity after the infectious period~\cite{Bjornstad2018}. In this subsection, for simplicity, we consider only permanent immunity. A new compartment called recovered ($R$) is included to hold the individuals who acquire permanent immunity. Then, the simplest model with this characteristic is susceptible-infected-recovered or the SIR model. A schematic representation of the SIR model is exhibited in Fig.~\ref{fig18}, where the new compartment is represented by the blue square. Now the interpretation of $\gamma$ is clearer and is named as removal or recovery rate, as well as the reciprocal $1/\gamma$ is the average infectious period. After the infectious period, the individuals acquire permanent immunity.

To introduce the ODE formulation, we consider a close population, i.e., without deaths, births, or migrations (demographics parameters). Also, we maintain the consideration that the population is homogeneous mixing. Then, the equations for the SIR model are
\begin{eqnarray}
\label{sir-s}
\frac{ds}{dt} &=& -\beta si, \\
\label{sir-i}
\frac{di}{dt} &=& \beta si - \gamma i, \\
\label{sir-r}
\frac{dr}{dt} &=& \gamma i,
\end{eqnarray}
{ similar to the previous models, the considered solutions for SIR model are restricted to $D = \{(s,i,r) \in [0,1]^3 \ |  \ s(t) + i(t) + r(t) = 1 \}$ and the parameters $s_0$, $i_0$, $r_0$, $\alpha$, $\beta$, and $\gamma$ are positive.}
 Due to the non-linearity term ($si$), these equations can not be solved explicitly. However, before we study the numerical solutions, let us study some dynamical properties of the model.

According to the SIS model, the basic reproductive number ($R_0$) indicates when an infectious agent is successful in starting an epidemic or unsuccessful. The phenomenon related to $R_0$ and the beginning of an epidemic situation is called the threshold phenomenon. Initially, the susceptible fraction needs to exceed the $\gamma/\beta$ fraction. This property is demonstrated by taking Eq.~(\ref{sir-i}) and rewrites as $di/dt=i\beta(s - \gamma/\beta)$. If $s>\gamma/\beta$,  then $di/dt>0$ and, consequently, the infected population increases. In this case, the basic reproduction number is $R_0 = \beta/\gamma$.

Another important phenomenon is the asymptotic state of the SIR model. To obtain this expression, we divide Eq.~(\ref{sir-s}) by Eq.~(\ref{sir-r}), which results in
\begin{equation}
\frac{ds}{dr}=-\frac{\beta s}{\gamma} \equiv - R_0 s.
\label{sir-eq1}
\end{equation}
Considering $s(0) = s_0$ and $r_0 = 0$, the solution is
\begin{equation}
s(t) = s_0 e^{-R_0 r(t)}.
\label{sir-eq2}
\end{equation}
This result shows that $s$ individuals decay exponentially as $r$ increases. The number of recovered increases with a delay time proportional to $1/\gamma$. The values of $s_0$ and $e^{-R_0 r(t)}$ are always positive, then the $s(t)$ remains above zero. 

\begin{figure}[!hbt]
	\centering 
	\includegraphics[scale=1.0]{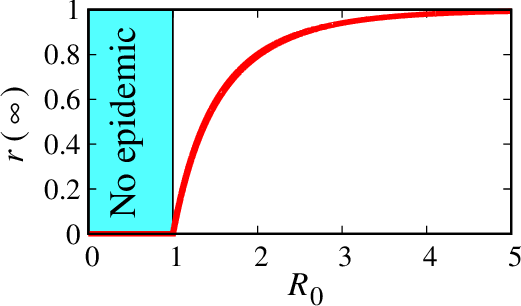} 
	\caption{Fraction of infected ($r(\infty)$) versus $R_0$ for $s_0=1$.}
	\label{fig19}
\end{figure}

The result obtained in Eq.~(\ref{sir-eq2}) leads to an important and counter-intuitive conclusion: the spread eventually stops due to the decline of infectious, not due to the complete absence of susceptibility. We consider $i=0$ in the final spread ($t \rightarrow \infty$) to obtain a complete description of this statement. By considering the constraint, $s+i+r=1$, we get $s=1-r$. Using the results from Eq.~(\ref{sir-eq2}), we obtain
\begin{equation}
1 - r(\infty) - s_0 e^{-R_0 r(\infty)} = 0,
\end{equation}
where $r(\infty)$ is the final fraction of recovered individuals, which is equal to the population fraction that gets infected. This is a transcendental equation, and the solution for $s(0)=1$ is obtained numerically, as shown in Fig.~\ref{fig19}. This result is very important once it corroborates the threshold phenomenon. In the SIS model, we discuss the threshold phenomenon showing that an epidemic occurs only when $R_0 > 1/s_0$. Once $s_0 = 1$, the epidemic situation occurs only for $R_0$. In the result exhibited in Fig.~\ref{fig19} for values below $R_0<1$, there are no recovered individuals, which signifies no epidemic. On the other hand, for $R_0>1$, the infection invades the population. The result shows that for $R_0 =2$, the fraction of individuals who get the disease is $\approx 0.80$.

\begin{figure}[!hbt]
	\centering 
	\includegraphics[scale=0.5]{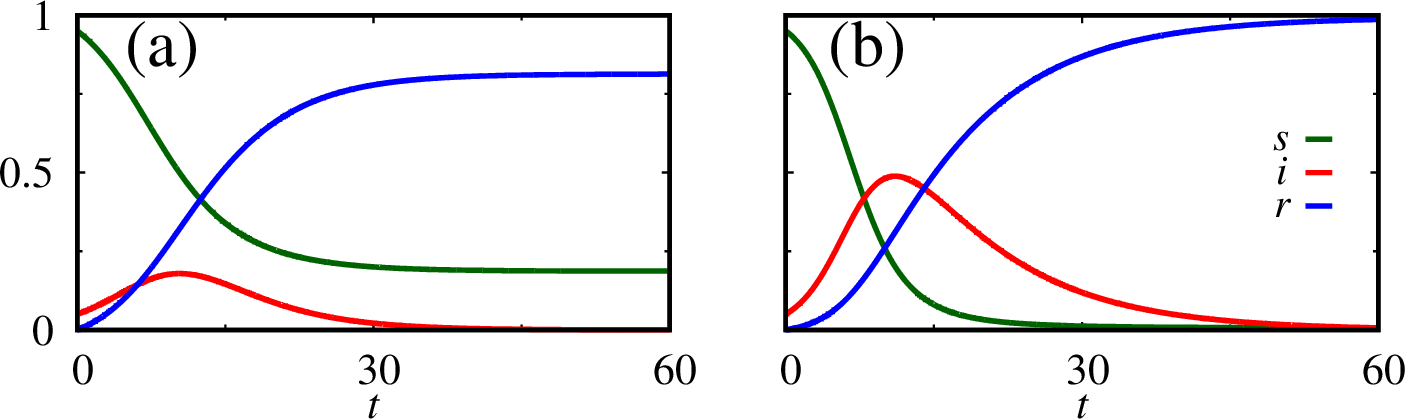}
	\caption{Solutions for SIR model.
	In the panel (a) $R_0 = 2$ ($\beta=0.5$ and $\gamma=0.25$).
	In the panel (b) $R_0 = 5$ ($\beta=0.5$ and $\gamma=0.1$). We consider $i_0 = 0.05$ and $r_0 = 0.0$.}
	\label{fig20}
\end{figure}

The numerical solution of Eqs.~(\ref{sir-s}), (\ref{sir-i}) and (\ref{sir-r}) are shown in Fig.~\ref{fig20}, where the green, red and blue lines correspond to $s$, $i$ and $r$, respectively. The panel (a) is for $R_0=2$. If  $R_0 =2$, $\approx 0.80$ of the population gets an infection. This result is displayed in Fig.~\ref{fig20}(a) in the recovered curve (green line). On the other hand, if $R_0>5$, approximately all individuals are infected, as we see in Fig.~\ref{fig19}. In this way, the panel \ref{fig20}(b)
displays the results for $R_0=5$, where it is possible to see approximately all individuals getting an infection. From these results, it is straightforward to see that one compartment runs out due to the filling of the other.

%%%%%%%%%%%%%%%%%%%%%%%%%%%%%%%%%%%%%%%%%%
\subsubsection{SIR fractional model}

The fractional extension of the SIR model is given by
\begin{eqnarray}
\label{sir-fractional-s}
\frac{d^\alpha s}{dt^\alpha} &=& -\beta si, \\
\label{sir-fractional-i}
\frac{d^\alpha i}{dt^\alpha} &=& \beta si - \gamma i, \\
\label{sir-fractional-r}
\frac{d^\alpha r}{dt^\alpha} &=& \gamma i,
\end{eqnarray}
{ where the considered solutions remain in $D = \{(s,i,r) \in [0,1]^3 \ |  \ s(t) + i(t) + r(t) = 1 \}$ with initial conditions and parameters positives.} The basic reproductive number is given by $R_0 = \beta/\gamma$. An expression type Eq.~(\ref{sir-eq2}) is not possible due to the appearance of the term $D^\alpha s / s$, in which the Laplace transformation diverges. Nevertheless, the threshold phenomenon remains valid. 

\begin{figure}[!hbt]
	\centering 
	\includegraphics[scale=0.5]{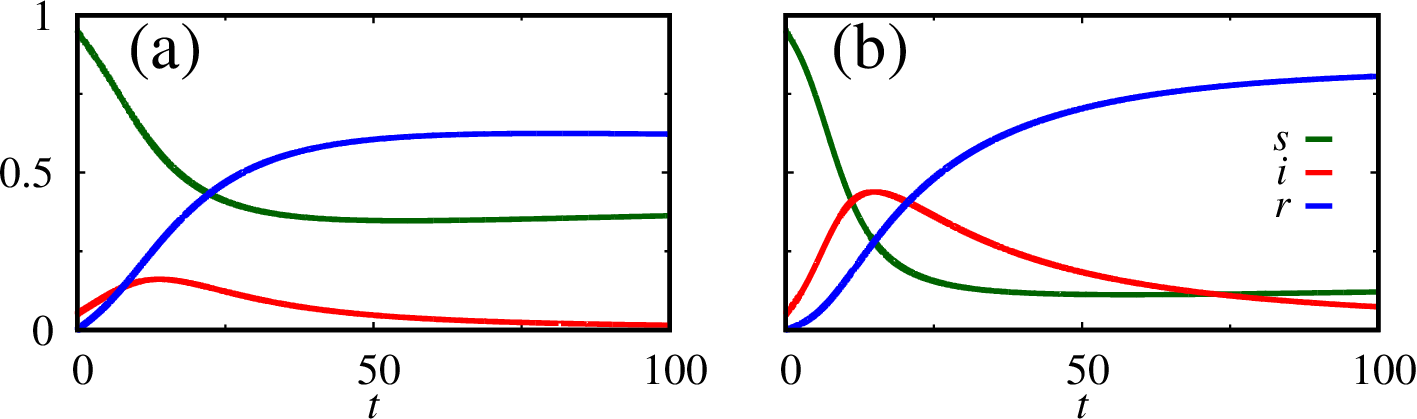}
	\caption{Solutions for fractional SIR model with $\alpha=0.8$.
	In the panel (a) $R_0 = 2$ ($\beta=0.5$ and $\gamma=0.25$).
	In the panel (b) $R_0 = 5$ ($\beta=0.5$ and $\gamma=0.1$).  We consider $i_0 = 0.05$ and $r_0 = 0.0$.}
	\label{fig21}
\end{figure}

Numerical solutions for Eqs.~(\ref{sir-fractional-s}), (\ref{sir-fractional-i}) and (\ref{sir-fractional-r}) are shown in Fig.~\ref{fig21}. The panel (a) is for $R_0 = 2$ ($\beta = 0.5$ and $\gamma=0.25$) and the panel (b) for $R_0 = 5$ ($\beta=0.5$ and $\gamma=0.1$). As observed in the previous models, the fractional effects impose a lag in the time to reach the steady solution, where in both cases are $i=0$. We do not obtain an expression to determine the steady value of the respective fractions; numerically, these values correspond to one obtained in the standard formulation.

\begin{figure}[!hbt]
	\centering
	\includegraphics[scale=1.0]{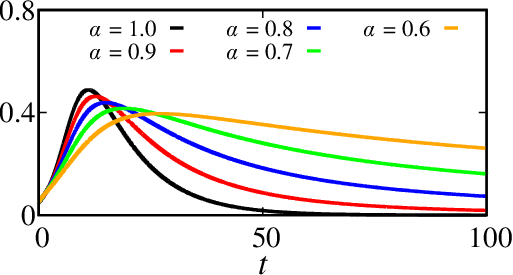} 
	\caption{Comparison between standard (black line) and fractional SIR 
	model for $\alpha=0.9$ (red line), $\alpha=0.8$ (blue line), $\alpha=0.7$ 
	(green line) and $\alpha=0.6$ (orange line). 
	We consider $R_0 = 5$ ($\beta=0.5$ and $\gamma=0.1$), $i_0 = 0.05$, and $r_0 = 0.0$.}
	\label{fig22}
\end{figure}

Considering $R_0 = 5$ ($\beta=0.5$ and $\gamma=0.1$), Fig.~\ref{fig22} displays solutions for $\alpha=1$ (black line), $\alpha=0.9$ (red line), $\alpha=0.8$ (blue line), $\alpha=0.7$ (green line) and $\alpha=0.6$ (orange line). The results exhibit the influence of flattening the infected curve; however, the area below the curve is practically the same. Despite the peak being smaller for small $\alpha$, the curve takes longer to reach the steady solution. 

%%%%%%%%%%%%%%%%%%%%%%%%%%%%%%%%%%%%%%%%%%
\subsubsection{SIR fractal model}

The fractal extension of the SIR model is similar to SI and SIS models, which is 
\begin{eqnarray}
\label{sir-fractal-s}
\frac{ds}{dt} &=& - \alpha t^{\alpha-1} \beta si, \\
\label{sir-fractal-i}
\frac{di}{dt} &=& \alpha t^{\alpha-1} \left(\beta si - \gamma i\right), \\
\label{sir-fractal-r}
\frac{dr}{dt} &=& \alpha t^{\alpha-1} \gamma i,
\end{eqnarray}
{ the analyzed solutions remain in the set $D = \{(s,i,r) \in [0,1]^3 \ |  \ s(t) + i(t) + r(t) = 1 \}$ with initial conditions and parameters positives.}
The basic properties, such as the constraint, remain unchanged. Assuming that $\alpha$ and $t$ are non-null, the division of Eq.~(\ref{sir-fractal-s}) by Eq.~(\ref{sir-fractal-r}) results in  Eq.~(\ref{sir-eq1}) with solution given by Eq.~(\ref{sir-eq2}). The only correction is in the unities.

\begin{figure}[!hbt]
	\centering 
	\includegraphics[scale=0.5]{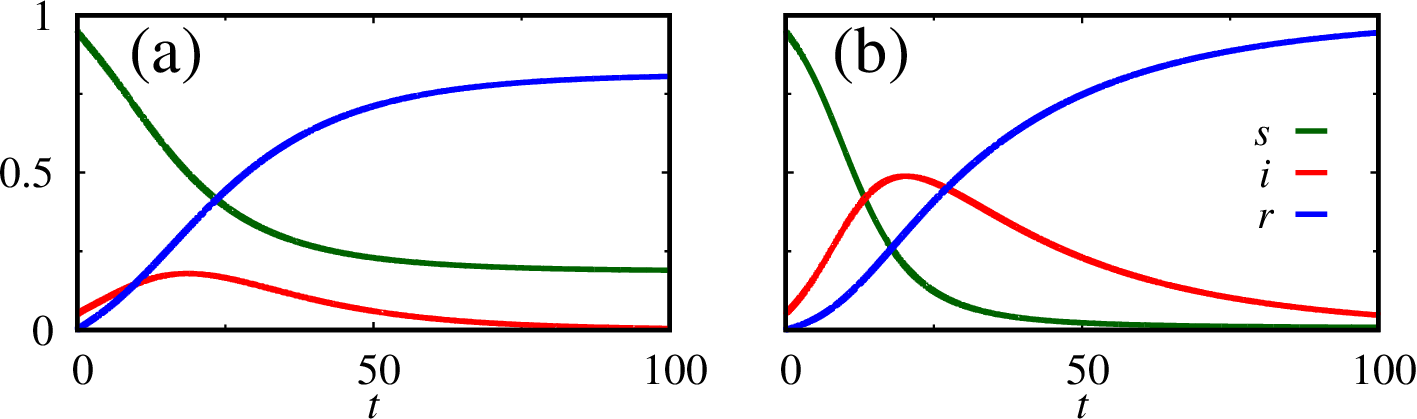}
	\caption{Solutions for fractal SIR model with $\alpha=0.8$.
	In the panel (a) $R_0 = 2$ ($\beta=0.5$ and $\gamma=0.25$).
	In the panel (b) $R_0 = 5$ ($\beta=0.5$ and $\gamma=0.1$). We consider $i_0 = 0.05$ and $r_0 = 0.0$.}
	\label{fig23}
\end{figure}

A numerical solution for the SIR fractal model with $\alpha=0.8$ is displayed in Fig.~\ref{fig23} in panels (a) and (b) for $R_0 = 2$ and $R_0 = 5$, respectively. As in the fractional case, including fractal operators with an order of less than 1 makes the spread of disease slower compared with the standard case.

\begin{figure}[!hbt]
	\centering
	\includegraphics[scale=1.0]{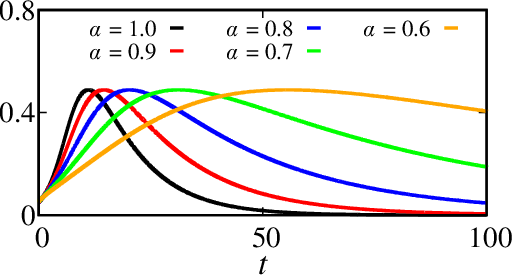} 
	\caption{Comparison between standard (black line) and fractal SIR 
	model for $\alpha=0.9$ (red line), $\alpha=0.8$ (blue line), $\alpha=0.7$ 
	(green line), and $\alpha=0.6$ (orange line). 
	We consider $R_0 = 5$ ($\beta=0.5$ and $\gamma=0.1$), $i_0 = 0.05$, and $r_0 = 0.0$.}
	\label{fig24}
\end{figure}

The comparison between the standard (black line) and fractal (colorful lines) formulation of the SIR model is displayed in Fig.~\ref{fig24} for $R_0 = 5$, for $\alpha=0.9$ (red line), $\alpha=0.8$ (blue line), $\alpha=0.7$ (green line), and $\alpha=0.6$ (orange line). In this formulation, the curve becomes wider, and the peak remains, practically, at the same level as when $\alpha=1$. As $\alpha$ decreases, the time to reach the steady solution increases.

\begin{figure}[!hbt]
	\centering
	\includegraphics[scale=1.0]{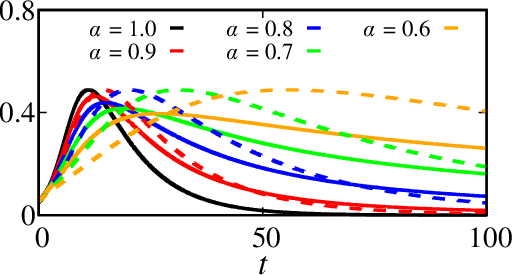} 
	\caption{Comparison between the three formulations for the SIR model: 
	black line standard model, continuous fractional, and dotted fractal approach.  
	In value, the constants are $\beta=0.5$, $\gamma=0.1$, $i_0 = 0.05$ and $r_0 = 0.0$.}
	\label{fig25}
\end{figure}

Now, with the goal of comparing the three formulations, Fig.~\ref{fig25} exhibits the solution for the standard (black line), fractional (colorful continuous), and fractal (colorful dotted lines) formulations. This result shows that the fractal formulation differs from the fractional one in the time to reach the steady solution and in the peak and wider of the infected curve.

%%%%%%%%%%%%%%%%%%%%%%%%%%%%%%%%%%%%%%%%%%%%%%%%%%%
\subsubsection{Application of SIR model}

\begin{figure}[!hbt]
	\centering
	\includegraphics[scale=1.0]{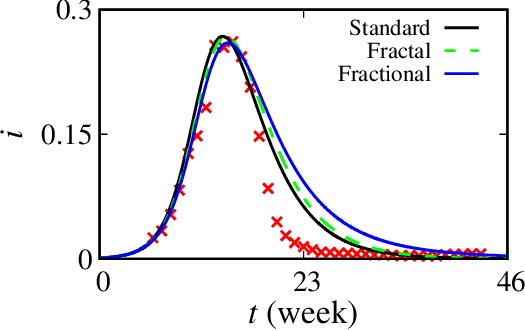} 
	\caption{Description of Influenza cases in three SIR formulations: standard (black 
	line), fractal (dotted green line, $\alpha=0.99$) and fractional (blue line, $\alpha=0.95$). We 
	consider $\beta=5.95$, $\gamma=2.17$ (1/week), $i_0 = 0.001$ and $R_0 = 2.7$.}
	\label{fig26}
\end{figure}

One application of the SIR model is to describe the influenza behaviour~\cite{Towers2011, Laguzet2015}. Considering the percentage of infected individuals by Influenza A (H1N1, H3N2, H3N2v, sub-typing not performed) from the Centers of Disease Control and Prevention (CDC)\footnote{Available on: https://www.cdc.gov/flu/weekly/index.htm Accessed on 4 Jul. 2023.}, we use the three formulations of SIR model to describe the curve. The result is displayed in Fig.~\ref{fig26}. The points are associated with the real data, while the lines are with the models. The black line is for the standard SIR, the green dotted line for the fractal SIR with $\alpha=0.99$, and the blue line for the fractional SIR with $\alpha=0.95$. We consider the parameters equal to $\beta=5.95$, $\gamma=2.17$ (1/week) and $i_0 = 0.001$. The data are described by the models until the peak; after that, the models lose accuracy. This occurs due to medical or social interventions, which are not considered in the models. However, at the beginning of the epidemic, the fractional formulation describes the data more accurately than the other ones.

%%%%%%%%%%%%%%%%%%%%%%%%%%%%%%%%%%%%%%%%%%%%%%%%%%%
\subsection{Susceptible-Infected-Exposed-Recovered (SEIR) model}
%%%%%%%%%%%%%%%%%%%%%%%%%%%%%%%%%%%%%%%%%%

\subsubsection{SEIR standard model}

\begin{figure}[!hbt]
	\centering 
	\includegraphics[scale=0.4]{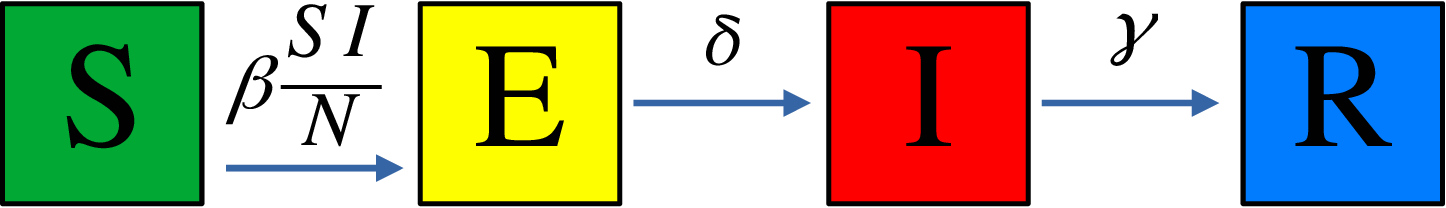} 
	\caption{Schematic representation of the SEIR model.}
	\label{fig27}
\end{figure}

For some diseases, such as COVID-19~\cite{He2020, IHME2020}, the infected individuals do not start to be infectious right after acquiring the illness. They are infected and spend an average time equal to $1/\delta$ to become infectious. This period is called latent~\cite{Gabrick2022}. The compartment that holds the individuals is the exposed (E). A schematic representation of the susceptible-infected-exposed-recovered (SEIR) model is displayed in Fig.~\ref{fig27}.

The  ODE description of the SEIR model is very similar to the SIR model, the difference is in a new equation that makes the flow between $S$ to $E$ and $E$ to $I$. The equations are
\begin{eqnarray}
\label{seir-s}
\frac{ds}{dt} &=& -\beta si, \\
\label{seir-e}
\frac{de}{dt} &=& \beta si - \delta e, \\
\label{seir-i}
\frac{di}{dt} &=& \delta e -\gamma i, \\
\label{seir-r}
\frac{dr}{dt} &=& \gamma i.
\end{eqnarray}
{ Similar to the previous models, we only consider positive solutions for SEIR model, such as $D = \{(s,e,i,r) \in [0,1]^4 \ |  \ s(t) + e(t) + i(t) + r(t) = 1 \}$ and $s_0$, $e_0$, $i_0$, $r_0$, $\beta$, $\delta$, and $\gamma \geq 0$.} In this way, Eq.~(\ref{seir-r}) can be replaced by the algebraic equation $r=1-s-e-i$. The equilibrium solutions of the SEIR model are the disease-free $(\widehat{s},\widehat{i}, \widehat{e}) = (1,0,0)$ and the endemic solution is $(s^{*}, e^{*}, i^{*}) = (R_0^{-1},0,0)$, where $r = 1 - s -e - i$ and $R_0 = \beta/\gamma$~\cite{Keeling2008}.

\begin{figure}[!hbt]
	\centering 
	\includegraphics[scale=1]{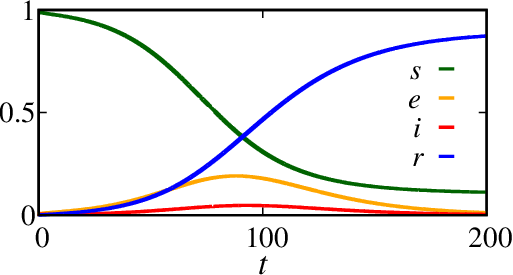}
	\caption{Solution for standard SEIR model.
	    We consider $\beta=0.5$, $\gamma=0.2$, $\delta=0.05$ ($R_0 = 2.5$) and $(s_0,e_0,i_0,r_0) = (0.99,0.005,0.005,0)$.}
	\label{fig28}
\end{figure}

Figure~\ref{fig28} exhibits a numerical solution for the standard formulation. We consider $\beta=0.5$, $\gamma=0.2$, $\delta=0.05$ ($R_0 = 2.5$) and $(s_0,e_0,i_0,r_0) = (0.99,0.005,0.005,0)$. At the beginning of propagation, the fraction $e$ start to increase while $s$ decrease. As $e$ evolves to $i$, the curve related to $i$ starts to increase, and after $1/\gamma$, the individuals evolve to $r$. The latent period is $1/\delta = 20$ and the infectious $1/\gamma=10$. As the latent period is greater than infectious, the individuals stay for more time in $e$, which is reflected by a higher amplitude of $e$ curve compared with $i$. However, after reaching the peak, the $e$ curve starts to decrease, and the $i$ also follows the same behavior. The disease-free is found only when $e=i=0$.

%%%%%%%%%%%%%%%%%%%%%%%%%%%%%%%%%%%%%%%%%%
\subsubsection{SEIR fractional model}

The SEIR model extended by fractional operators is governed by the following equations:
\begin{eqnarray}
\label{seir-fractional-s}
\frac{d^\alpha s}{dt^\alpha} &=& -\beta si, \\
\label{seir-fractional-e}
\frac{d^\alpha e}{dt^\alpha} &=& \beta si - \delta e, \\
\label{seir-fractional-i}
\frac{d^\alpha i}{dt^\alpha} &=& \delta e -\gamma i, \\
\label{seir-fractional-r}
\frac{d^\alpha r}{dt^\alpha} &=& \gamma i,
\end{eqnarray}
{ the solutions are restricted to $D = \{(s,e,i,r) \in [0,1]^4 \ |  \ s(t) + e(t) + i(t) + r(t) = 1 \}$ and $s_0$, $e_0$, $i_0$, $r_0$, $\alpha$, $\beta$, $\delta$, and $\gamma \geq 0$.} Eq.~(\ref{seir-fractional-r}) can be replaced by the algebraic equation $r = 1 - s - e -i$. The basic reproduction number and equilibrium solutions remain equal to the standard case; the only difference is the unit correction. This is because, in the Caputo sense, the derivative of a constant remains zero.

\begin{figure}[!hbt]
	\centering 
	\includegraphics[scale=1]{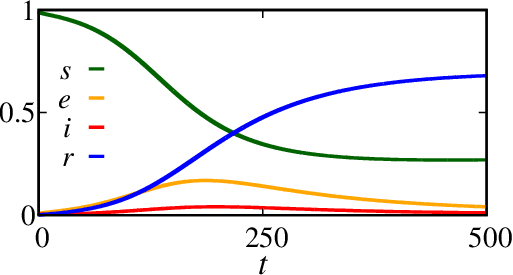}
	\caption{Solution for fractional SEIR model with $\alpha=0.8$.
	    We consider $\beta=0.5$, $\gamma=0.2$, $\delta=0.05$ ($R_0 = 2.5$) and $(s_0,e_0,i_0,r_0) = (0.99,0.005,0.005,0)$.}
	\label{fig29}
\end{figure}

Figure~\ref{fig29} shows a numerical solution for the fractional SEIR model for $\alpha=0.8$, $\beta=0.5$, $\gamma=0.2$, $\delta=0.05$ ($R_0 = 2.5$) and $(s_0,e_0,i_0,r_0) = (0.99,0.005,0.005,0)$. As observed in the previous results, the effects of fractional operators made the transition between compartments slow compared with the standard case. The solutions spend more time to reach a steady solution, which, in this case, is disease-free. 

\begin{figure}[!hbt]
	\centering
	\includegraphics[scale=1.0]{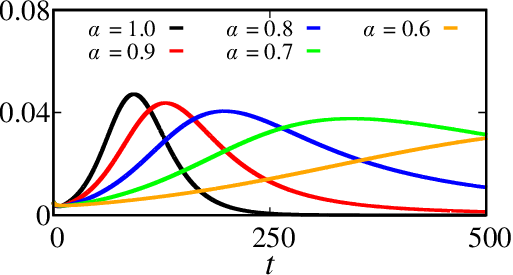} 
	\caption{Comparison between standard (black line) and fractional SEIR 
	model for $\alpha=0.9$ (red line), $\alpha=0.8$ (blue line), $\alpha=0.7$ 
	(green line), and $\alpha=0.6$ (orange line). 
	We consider $\beta=0.5$, $\gamma=0.2$, $\delta=0.05$ ($R_0 = 2.5$) and $(s_0,e_0,i_0,r_0) = (0.99,0.005,0.005,0)$.}
	\label{fig30}
\end{figure}

A comparison between the fractional and standard formulations for the $i$ variable is exhibited in Fig.~\ref{fig30}, where $\alpha=1$ (black line), $\alpha=0.9$ (red line), $\alpha=0.8$ (blue line), $\alpha=0.7$ (green line) and $\alpha=0.6$ (orange line). The results show that the peak of the infected curve decay and the wider curve increase as a function of $\alpha$ decreasing. The opening of the infected curve is related to the constraint, which is always preserved.  

%%%%%%%%%%%%%%%%%%%%%%%%%%%%%%%%%%%%%%%%%%
\subsubsection{SEIR fractal model}

A fractal extension of the SEIR model is described by the equations:
\begin{eqnarray}
\label{seir-fractal-s}
\frac{ds}{dt} &=& - \alpha t^{\alpha-1} \beta si, \\
\label{seir-fractal-e}
\frac{de}{dt} &=& \alpha t^{\alpha-1}  (\beta si - \delta e), \\
\label{seir-fractal-i}
\frac{di}{dt} &=& \alpha t^{\alpha-1}  (\delta e -\gamma i), \\
\label{seir-fractal-r}
\frac{dr}{dt} &=& \alpha t^{\alpha-1}  \gamma i,
\end{eqnarray}
{ we consider the solutions into $D = \{(s,e,i,r) \in [0,1]^4 \ |  \ s(t) + e(t) + i(t) + r(t) = 1 \}$ and positive parameters and initial conditions.} The basic reproduction number remains unchanged as the equilibrium solution differs only in the unities correction.

\begin{figure}[!hbt]
	\centering 
	\includegraphics[scale=1]{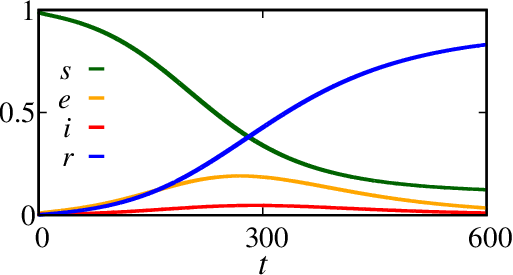}
	\caption{Solution for fractal SEIR model with $\alpha=0.8$.
	    We consider $\beta=0.5$, $\gamma=0.2$, $\delta=0.05$ ($R_0 = 2.5$) and $(s_0,e_0,i_0,r_0) = (0.99,0.005,0.005,0)$.}
	\label{fig31}
\end{figure}

Considering the parameters equal to $\alpha=0.8$, $\beta=0.5$, $\gamma=0.2$, $\delta=0.05$ ($R_0 = 2.5$) and $(s_0,e_0,i_0,r_0) = (0.99,0.005,0.005,0)$, a numerical solution is shown in Fig.~\ref{fig31}. The global aspect of the solution remains equal to both previous formulations. The difference is in the time to reach a steady solution. Now, due to the effects of the fractal operator, the solutions take a long time to reach a disease-free equilibrium. 

\begin{figure}[!hbt]
	\centering
	\includegraphics[scale=1.0]{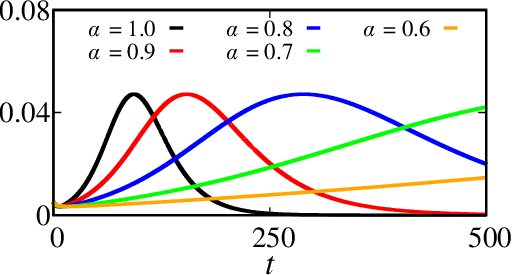} 
	\caption{Comparison between standard (black line) and fractal SEIR 
	model for $\alpha=0.9$ (red line), $\alpha=0.8$ (blue line), $\alpha=0.7$ 
	(green line), and $\alpha=0.6$ (orange line). 
	We consider $\beta=0.5$, $\gamma=0.2$, $\delta=0.05$ ($R_0 = 2.5$) and $(s_0,e_0,i_0,r_0) = (0.99,0.005,0.005,0)$.}
	\label{fig32}
\end{figure}

A comparison between the fractal (dotted lines) and standard (black line) SEIR model is exhibited in Fig.~\ref{fig32} for the $i$ state. As we note from the result, how less is $\alpha$ more time the curve spends to reach the peak value. Consequently, more time the system leads to reaching the equilibrium state. 

\begin{figure}[!hbt]
	\centering
	\includegraphics[scale=1.0]{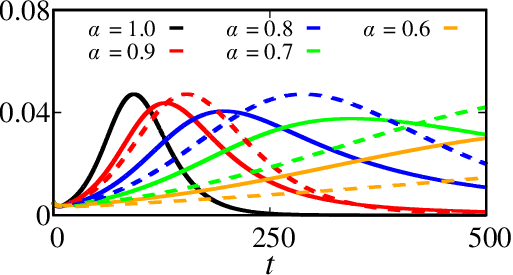} 
	\caption{Comparison between the three formulations for the SEIR model: 
	black line standard model, continuous fractional, and dotted fractal approach.  
	We consider $\beta=0.5$, $\gamma=0.2$, $\delta=0.05$ ($R_0 = 2.5$) and $(s_0,e_0,i_0,r_0) = (0.99,0.005,0.005,0)$.}
	\label{fig33}
\end{figure}

As a final comparison among the three formulations, Fig.~\ref{fig33} displays the result for standard (black line), fractional (continuous colorful), and fractal (dotted colorful) SEIR formulations. Now, the difference in the infected curve for the three formulations is clear. In standard one, the curve increases as an exponential, then it reaches the peak and decreases until $i=0$. In the fractional (continuous lines), the infected curve spends more time reaching the peak value and decreases with slower velocity. On the other hand, in the fractal, the curve increases and decreases at an approximate rate. However, this rate is less compared with the standard. The peaks for the three formulations have different values. 

%%%%%%%%%%%%%%%%%%%%%%%%%%%%%%%%%%%%%%%%%%
\subsubsection{Application of SEIR model}

\begin{figure}[!hbt]
	\centering
	\includegraphics[scale=1.0]{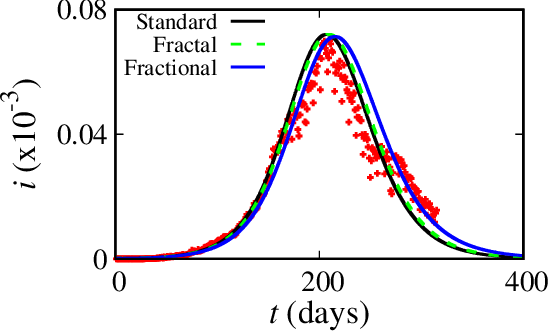} 
	\caption{Description of COVID-19 cases in three SEIR formulations: standard (black 
	line), fractal (dotted green line, $\alpha=0.998$) and fractional (blue line, $\alpha=0.988$). We 
	consider $\beta=0.2$, $\gamma=0.1$, $\delta=0.09$ (1/day) and $i_0 = 0.0002$.}
	\label{fig34}
\end{figure}

One application of the SEIR model that gained much attention in the last years is in modeling COVID-19. Here, we consider the data from India from 2020-04-10 to 2020-12-31 (265 days)\footnote{Available on: https://covid19.who.int/data Accessed on: 06 Jul. 2023}. Figure~\ref{fig34} shows the plot of data by the red points and the adjustment given by the three formulations of the SEIR model: standard (black line), fractal with $\alpha=0.998$ (dotted green line), and fractional with $\alpha=0.988$ (blue line). Our parameters are $\beta=0.2$, $\gamma=0.1$, $\delta=0.09$ and $i_0 = 0.0002$ in day units. To maintain fractions of infected, we consider the available data and divide by the population size in India in 2020, which was $1.396 \times 10^9$. { The three frameworks describe the data with great accuracy. Nonetheless, the standard formulation fit well the points when compared with the other ones. }

%%%%%%%%%%%%%%%%%%%%%%%%%%%%%%%%%%%%%%%%%%%%%%%%%%%
\section{Conclusion}

In this Chapter, we consider four epidemic models: SI, SIS, SIR, and SEIR. We introduce some basic concepts related to epidemiological models. Firstly, we start discussing all the models in the standard approach, i.e., considering the formulations based on ordinary differential equations. After that, we introduce the extensions of the respective models by non-integer operators. We consider two extensions given by fractional and fractal operators. For the fractional one, we utilize the Caputo definition and the Hausdorff derivative for the fractal. As a global aspect, our results suggest that the models reach steady solutions; however, the time and the form to reach them depend on the formulation. In the standard and fractal, which are Markovian, the time to reach the steady solution is earlier compared with the fractional, which are non-Markovian. This time is amplified as a function of the decrease in derivative order. However, because the population remains constant when the order of the operator is decreased, the infected curve is wider. In addition, the peak of the infected curve depends on the formulation.

For the four models, we consider real data. For the SI, we consider data from AIDS, and our results suggest that the fractal order describes the data with more accuracy. In the SIS model, for the data from syphilis, the three formulations describe the data behavior with good accuracy. Concerning the SIR model, we use the data of Influenza. In this case, the fractional formulation describes the data more accurately than the other two models until the peak point. As our last model, we apply the SEIR to the COVID-19 data from India. In this case, the model that describes the data is the standard one. The other two describe the accuracy for orders near 1. 

All in all, our results suggest that the different approaches considered to study the epidemiological models may bring improvements or not. The non-integer formulations described better the data in some cases, but in others, they failed. 
{ In addition, it is important to note that the fractional models spend more computational time than standard or fractal, and in general they are more difficult to analyze. Comparing the fractal and the standard model, both spend approximately the same computational cost. The only problem is that the fractal formulation becomes non-autonomous, which brings some difficulty to analyze the stability. 
The choice of a formulation depends on the type of data and the problem to be investigated. It is important to note, that in this Chapter we present the models in their simplest form, i.e., without any additional terms. Future works can be conducted by including new compartments and new equations.}

%%%%%%%%%%%%%%%%%%%%%%%%%%%%%%%%%%%%%%%%%%%%%%%%%%%
\section*{Acknnowledgements}
The authors thank the financial support from the Brazilian Federal Agencies (CNPq); CAPES; Funda\-\c c\~ao A\-rauc\'aria. S\~ao Paulo Research Foundation (FAPESP 2022/13761-9). E.K.L. acknowledges the support of the CNPq (Grant No. 301715/2022-0). E.C.G. received partial financial support from Coordenação de Aperfeiçoamento de Pessoal de Nível Superior - Brasil (CAP ES) - Finance Code 88881.846051/2023-01. We would like to thank 105 Group Science (www.105groupscience.com).

%%%%%%%%%%%%%%%%%%%%%%%%%%%%%%%%%%%%%%%%%%%%%%%%%%%
\newpage
\bibliographystyle{elsarticle-num}
\bibliography{references}
\end{document}